\documentclass[a4paper,11pt]{article}
\usepackage{jheppubMOD} 
\usepackage{lipsum}
\usepackage{graphicx}  
\usepackage{multirow}
\usepackage[x11names,dvipsnames]{xcolor}

\hypersetup{
    colorlinks,
    linkcolor={Blue},
    citecolor={Blue},
    urlcolor={Blue}
}

\usepackage{lmodern}
\usepackage{amsmath,amssymb}
\usepackage{mathtools}
\usepackage[numbers,sort&compress]{natbib}

\usepackage{tikz}
\usepackage{tikz-3dplot}
\usepackage{pgfplots,pgfplotstable}
\usetikzlibrary{pgfplots.groupplots}
\usetikzlibrary{decorations.markings,decorations.pathmorphing}
\usetikzlibrary{shapes}
\usetikzlibrary{shapes.misc}
\usetikzlibrary{shapes.geometric}
\usetikzlibrary{shapes.misc}
\usetikzlibrary{patterns}
\usetikzlibrary{positioning}
\usetikzlibrary{3d,calc}
\usetikzlibrary{trees}
\usetikzlibrary{arrows}
\usetikzlibrary{through}
\usetikzlibrary{plotmarks}

\usepgfplotslibrary{fillbetween}
\pgfplotsset{compat=newest}

\usepackage{shellesc}
\usetikzlibrary{external}
\tikzsetexternalprefix{figures/}
\usepgfplotslibrary{external} 
\pgfplotsset{table/search path={ResFiles/},}

\usepackage{stackrel}
\usepackage{bm}

\usepackage[compat=1.1.0]{tikz-feynman}

\usepackage{cleveref}
\usepackage{ifthen}
\usepackage{xspace}
\usepackage[shortlabels]{enumitem}

\usepackage[left=2cm,right=2cm,top=2cm,bottom=2cm]{geometry}

\usepgfplotslibrary{external}
\usetikzlibrary{external}
\pgfrealjobname{omegato3pi}

\newcommand{\be}{\begin{equation}}
\newcommand{\ee}{\end{equation}}
\newcommand{\bea}{\begin{eqnarray}}
\newcommand{\eea}{\end{eqnarray}}
\newcommand{\bef}{\begin{figure}}
\newcommand{\eef}{\end{figure}}
\newcommand{\bce}{\begin{center}}
\newcommand{\ece}{\end{center}}

\newcommand{\bD}{\bar D}

\newcommand{\ket}[1]{\left\lvert #1 \right\rangle}

\newcommand{\myPH}[3]{\vphantom{#1}\vphantom{#2}\vphantom{#3}}
\newcommand{\braketO}[3]{%
\left\langle #1 \myPH{#1}{#2}{#3} \right.%
\left\lvert #2 \myPH{#1}{#2}{#3} \right\rvert %
\left. #3 \myPH{#1}{#2}{#3} \right\rangle}

\newcommand{\tC}{\widetilde{C}}

\begin{document}

\preprint{JLAB-THY-21-3321}

\title{\boldmath $D\bar{D}^*$ scattering and $\chi_{c1}(3872)$ in nuclear matter}

\newcommand{\jlab}{Theory Center,
Thomas  Jefferson  National  Accelerator  Facility,
Newport  News,  VA  23606,  USA}
\newcommand{\ific}{Instituto de F\'isica Corpuscular (IFIC),
             Centro Mixto CSIC-Universidad de Valencia,
             Institutos de Investigaci\'on de Paterna,
             Aptd. 22085, E-46071 Valencia, Spain}
\newcommand{\ice}{Institute of Space Sciences (ICE, CSIC), Campus UAB,  Carrer de Can Magrans, 08193 Barcelona, Spain}
\newcommand{\ieec}{Institut d'Estudis Espacials de Catalunya (IEEC), 08034 Barcelona, Spain}
\newcommand{\fias}{Frankfurt Institute for Advanced Studies, Ruth-Moufang-Str. 1, 60438 Frankfurt am Main, Germany}
\newcommand{\stv}{Faculty  of  Science  and  Technology,  University  of  Stavanger,  4036  Stavanger,  Norway}

\author[a]{M.~Albaladejo}
\author[b]{J.~Nieves}
\author[c,d,e,f]{L.~Tolos}

\affiliation[a]{\jlab}
\affiliation[b]{\ific}
\affiliation[c]{\ice}
\affiliation[d]{\ieec}
\affiliation[e]{\fias}
\affiliation[f]{\stv}

\date{\today}

\abstract{We study the behaviour of the $\chi_{c1}(3872)$, also known as  $X(3872)$, in dense nuclear matter. We begin from a picture in vacuum of the $X(3872)$ as a purely molecular $(D \bar D^*-c.c.)$ state, generated as a bound state from a  heavy-quark symmetry leading-order interaction between the charmed mesons, and analyze the $D \bar D^*$ scattering $T-$matrix ($T_{D \bD^*}$) inside of the medium. Next, we consider also mixed-molecular scenarios and, in all cases, we determine the corresponding $X(3872)$ spectral function and the $D\bD^*$ amplitude, with the mesons embedded in the dense environment. We find important nuclear corrections for $T_{D \bD^*}$ and the pole position of the resonance, and discuss the dependence of these results on the  $D\bD^*$ molecular component in the $X(3872)$ wave-function. These predictions could be tested in the finite-density regime that can be accessed in the future CBM and PANDA experiments at FAIR.}
\frenchspacing
\toccontinuoustrue
\maketitle

\section{Introduction}

The new quarkonium revolution started in 2003 with the discovery of the $X(3872)$ (recently renamed as $\chi_{c1}(3872)$ \cite{Zyla:2020zbs}). It was firstly observed in $B^{\pm} \rightarrow K^{\pm} \pi^+ \pi^- J/\psi$ decays by Belle \cite{Choi:2003ue}, and subsequently confirmed by BaBar \cite{Aubert:2008gu}, CDF \cite{Acosta:2003zx,Abulencia:2006ma,Aaltonen:2009vj}, D$\varnothing$ \cite{Abazov:2004kp}, LHCb \cite{Aaij:2011sn,Aaij:2013zoa} and CMS \cite{Chatrchyan:2013cld}. The spin-parity quantum numbers $J^P=1^{++}$ were extracted at the 8$\sigma$ level in 2013 from the high-statistic measurements of the two-pion mode performed in the LHCb experiment \cite{Aaij:2015eva}. A distinctive feature of the $X(3872)$ is that the $\rho J/\psi$ and $\omega J/\psi$  branching fractions are similar. This points out to an isospin symmetry violation~\cite{Thomas:2008ja}, which together with the large disparity between $\omega$ and $\rho$ meson widths provides a natural explanation to the observed  $\rho J/\psi$ to $\omega J/\psi$ decay ratio \cite{Gamermann:2009fv, Gamermann:2009uq}.

The $X(3872)$ is one of the most studied exotic mesons with a $c \bar c$ content. This state lies extremely close to the $D^0 \bar D^{*0}$ threshold, and its (Breit-Wigner) width has been recently measured as $\Gamma = 1.39(24)(10)\ \text{MeV}$ \cite{Aaij:2020qga} or $\Gamma = 0.96(19)(21)\ \text{MeV}$ \cite{Aaij:2020xjx} in two different works by the LHCb collaboration. It can be produced via weak decays of $B$-mesons, that include two- (referred as $\rho J/\psi$ as it originates from $\rho$) and three-pion  (named as $\omega J/\psi$ as it comes from $\omega$) modes, or $\Lambda_b-$baryons,  as well as in charmonia radiative decays and through lepto- or photo-production. In addition, exhaustive sensitivity studies for width and line-shape measurements of the $X(3872)$ have been carried out for the reaction $p\bar{p} \to J/\psi \rho^0$ with the PANDA experiment at FAIR \cite{PANDA:2018zjt}, and the possibilities of $X(3872)$ photo-production off the nucleon have also been studied~\cite{Albaladejo:2020tzt}.

In spite of all this experimental progress, the nature of the $X(3872)$ is still elusive. From the point of view of constituent quark models, the most natural possibility for the $X(3872)$ is a $2\, ^3P_1$ $c \bar c$ charmonium configuration,\footnote{A heavy quark-antiquark bound state, characterized by the radial number $n$, the orbital angular momentum $L$, the spin $s$ and the total angular momentum $J$, is denoted by $n\, ^{2s+1}L_J$. Parity and charge conjugation are given by $P = (-1)^{L+1}$ and ${\cal C} = (-1)^{L+s}$, respectively.} {\it i.e.}, the $\chi_{c1}(2P)$ state. However, the quark model calculations give a value for the mass of this state higher than the experimental one (see, for example, Refs.~\cite{Badalian:1999fe,Barnes:2003vb,Barnes:2005pb}). Moreover, the isospin symmetry violation is difficult to explain using a simple $c \bar c$ model. Thus, new interpretations have been put forward. On the one hand, this state might be interpreted as a compact diquark and antidiquark (tetraquark) state  \cite{Maiani:2004vq,Ebert:2005nc,Matheus:2006xi}.  On the other hand, this state could be an example of a loosely bound hadron molecule (see, for example, Refs.~\cite{Tornqvist:2004qy,Wong:2003xk,Thomas:2008ja,Gamermann:2009fv, Gamermann:2009uq,Wang:2013daa}). The vicinity to the $D^0 \bar D^{*0}$ threshold and the large decay rate to $D^0 \bar D^{*0}$  together with a natural explanation of the isospin symmetry violation have made this interpretation quite popular. Also, other interpretations include hadrocharmonium \cite{Dubynskiy:2008mq}, a mixture between charmonium and exotic molecular states \cite{Matheus:2009vq, Ortega:2009hj,Cincioglu:2016fkm} or some relation with  a  $X$  atom,  which  is  a $D^{\pm} D^{*\mp}$ composite  system  with  positive  charge-conjugation and a mass of $\sim 3880$ MeV, formed mainly due to the Coulomb force \cite{Zhang:2020mpi}. For a detailed review of the present situation, we refer the reader to the recent reviews \cite{Esposito:2014rxa,Lebed:2016hpi,Esposito:2016noz,Guo:2017jvc,Olsen:2017bmm,Kalashnikova:2018vkv,Cerri:2018ypt, Brambilla:2019esw} and references therein. There are also several Lattice QCD simulations on the subject~\cite{Chiu:2006hd,Bali:2011dc,Bali:2011rd,Liu:2012ze,Yang:2012mya,Mohler:2012na,DeTar:2012xk,Bali:2012ua,Prelovsek:2013cra,Lee:2014uta,Padmanath:2015era,Cheung:2016bym}.

Most of these interpretations are based on the analysis of the charmonium spectrum and the comparison with the branching ratios for two- and three-body decays. However, the production of exotic charmonium in $pp$ reactions or relativistic heavy-ion collisions (HiCs)  has become a matter of recent interest as the production yield of these exotic hadrons could reflect their internal structure. 

The high prompt production cross section of the $X(3872)$ measured for $pp$ at CDF \cite{Abulencia:2006ma} and in CMS \cite{Chatrchyan:2013cld} has cast doubts on its possible interpretation as a $D^0 \bar D^{*0}$ molecule, since it was argued that the production of a weakly bound state should be strongly suppressed in high-energy collisions~\cite{Bignamini:2009sk}. However, this finding has been put into question in Ref.~\cite{Albaladejo:2017blx}, showing that the estimates for the cross sections using the molecular approach are consistent with the CDF and CMS measurements by an adequate election of the ultraviolet (UV) cutoff \cite{Albaladejo:2017blx}, a statement that has been in turn criticized in  Ref.~\cite{Esposito:2017qef}. Also, Ref.~\cite{Wang:2017gay}
questioned the production mechanism of the $X(3872)$ shown in Ref.~\cite{Bignamini:2009sk}, while conjecturing a new mechanism. The controversy has still continued in  Ref.~\cite{Braaten:2018eov}. In this latter work, it is shown that the prompt $X(3872)$ cross section at hadron colliders is consistent with those experimentally observed at CDF and CMS. This is concluded thanks to the derivation of a relation between the prompt $X(3872)$ cross section and that of a charm-meson pair, taking into account the threshold enhancement from the $X(3872)$ resonance. More recently, the production rates of promptly produced $X(3872)$ relative to the $\psi(2S)$ as a function of the final state particle multiplicity, obtained recently at LHCb, are explained within a comover interaction model if the $X(3872)$ is a tetraquark \cite{Esposito:2020ywk}. However, this result is again questioned in Ref.~\cite{Braaten:2020iqw} as it is argued that the breakup cross section are not well approximated by a geometric cross section inversely proportional to the binding energy of $X(3872)$, as assumed in Ref.~\cite{Esposito:2020ywk}. As a consequence, a simple modification of the comover model will give excellent fits to the LHCb data using parameters consistent with $X(3872)$ being a loosely bound charm-meson molecule. Thus, there is still an ongoing debate on the nature of the $X(3872)$ coming from the analysis of $pp$ collisions.

Another possible way to gain some insight about the nature of $X(3872)$ is to analyze its behavior for the extreme conditions present in HiCs at RHIC and LHC energies. The ExHIC Collaboration \cite{Cho:2010db,Cho:2011ew,Cho:2017dcy} has shown that, within the coalescence model, the molecular structure for the $X(3872)$ implies a production yield much larger than for the tetraquark configuration, in particular if one also takes into account the evolution in the hadronic phase \cite{Cho:2013rpa,Abreu:2016qci}. This is due to the fact that molecules are bigger than tetraquarks and, hence, the production and absorption cross sections in HiCs are expected to be larger. This was actually shown in Ref.~\cite{Abreu:2016qci}, where the time evolution of the $X(3872)$ abundance in the hot hadron gas was obtained, based on all the possible hadronic reactions for the production of $X(3782)$ of Ref.~\cite{Torres:2014fxa,Abreu:2016qci}. More recently the nature of $X(3872)$ in HiCs has been studied not only within instantaneous coalescence models \cite{Fontoura:2019opw,Zhang:2020dwn}, but also using a statistical hadronization model \cite{Andronic:2019wva} or by means of a thermal-rate equation approach \cite{Wu:2020zbx}. In those studies it is advocated that the quantitative description for a series of standard HiC observables, such as particle yields or the transverse spectra, might shed some light in the nature of the $X(3872)$.

The studies of the production of $X(3872)$ however do not  consider the possible in-medium modification of the hot hadronic phase. Only recently the behaviour of $X(3872)$ in a finite-temperature pion bath has been studied assuming this resonance to be a molecular state generated by the interaction of $D \bar D^* +c.c.$  pairs and associated coupled channels \cite{Cleven:2019cre}. The $X(3872)$ develops a substantial width, of the order of a few tens of MeV, within a hot pionic bath at temperatures 100-150 MeV,  whereas its nominal mass moves above the $D \bar D^*$ threshold. 

In the present work we address the behaviour of $X(3872)$ in a nuclear environment, with the objective of  analyzing the finite-density regime that can be accessed in HiCs and the future experiments at FAIR. An early study of the  $D_{s0}^*(2317)$ and  theorized $X(3700)$ scalar  mesons in a nuclear medium was performed in Ref.~\cite{Molina:2008nh}, already showing that the experimental analysis of the properties of those mesons is a valuable test of the nature of the open and hidden charm scalar resonances. More recently, in Ref.~\cite{Azizi:2017ubq} the in-medium mass shift of the $X(3872)$ was obtained using QCD sum rules, revealing that the mass of the resonance is considerably affected by the nuclear matter. 

We begin here from a picture of the $X(3872)$ as a molecular $D \bar D^*$ +c.c. state, generated as a purely molecular bound state from the leading-order interaction of the $D$ and $\bar D^*$ mesons, which is constrained by heavy-quark spin symmetry (HQSS) \cite{AlFiky:2005jd, HidalgoDuque:2012pq, Guo:2013sya, Albaladejo:2015dsa}. HQSS predicts that all types of spin interactions vanish for infinitely massive quarks, that is, the dynamics is unchanged under arbitrary transformations in the spin of the heavy quark. As a consequence, open charm pseudoscalar and vector mesons become degenerate in the infinite mass limit. We then implement the changes of the $D$ and $\bar D^*$ propagators in nuclear matter in order to obtain the in-medium $X(3872)$ scattering amplitude and the corresponding $X(3872)$ spectral function. Later on, we consider generalizations of the $D\bD^*$ interaction, allowing for scenarios in which the $X(3872)$ is not a purely molecular state, {\it i.e.}, it can be a compact state, and we also study mixed scenarios. In this way, we extract the modification on the mass and the width of $X(3872)$ in nuclear matter for different scenarios, in view of the forthcoming results on charmed particles in HiCs at CBM in FAIR \cite{Rapp:2011zz,Tolos:2013gta}. Moreover, the present study will be also of interest for PANDA, since it is expected that the  $X(3872)$ will strongly couple to the $\bar p p$ channel~\cite{PANDA:2018zjt}, and therefore, this resonance can be produced also in $\bar p A$ collisions~\cite{Larionov:2015nea}. Actually colliding antiprotons on nuclei with PANDA would allow the $A-$dependence of the production of $\psi(2S)$ and $X(3872)$ near threshold to be compared. This may, after appropriate theoretical study, provide a good way to expose an extended $D^*\bar D$ component of the $X(3872)$ state function~\cite{Lutz:2015ejy}. 

This work is organized as follows. In Sec.~\ref{sec:Tmatrix} we present the $D \bar D^*$ scattering amplitude and the $X(3872)$ in vacuum and in isospin-symmetric nuclear matter, while showing the open-charm ground-state spectral functions in matter. In Sec.~\ref{sec:self-energy} we determine the $X(3872)$ self-energy both in vacuum and in nuclear matter, while connecting the self-energy to the $D \bar D^*$ scattering amplitude. We finish by presenting our results in Sec.~\ref{sec:results}, and conclusions and future outlook in Sec.~\ref{sec:conclusions}. Finally in Appendix~\ref{app}, we give some details on the approximation used to extend the  nuclear medium  $D \bar D^*$ $T-$matrix  to the complex plane, allowing for the search of poles reported in  Sec.~\ref{sec:results}.

\section[\boldmath $T$-matrix formalism]{\boldmath $T$-matrix formalism}
\label{sec:Tmatrix}
\subsection[$D\bar{D}^*$ scattering amplitude and $X(3872)$]{\boldmath $D\bar{D}^*$ scattering amplitude and $X(3872)$}
To study the $X(3872)$ as a molecular state in the $D\bar{D}^*$  $I^G(J^{P{\cal C}}) = 0^+(1^{++})$ channel, we start by considering the interaction in the particle basis:
\begin{equation}
\{D^0 \bar{D}^{* 0}, D^{* 0}\bar{D}^0, D^+\bar{D}^{* -}, D^{* +}D^-\}~. \label{eq:particlebasis}
\end{equation}
The unitary $T$-matrix for this basis is written as:
\begin{equation}
T^{-1}(s) = V^{-1}(s) - \mathcal{G}(s)~, 
\end{equation}
with $\sqrt{s}$ the energy of any of the pairs in the center of mass (c.m.) frame, and  the $V$ and $\mathcal{G}$ matrices are constructed out of the interaction potential and the two-meson loop functions, respectively. From the leading order HQSS-based Lagrangian, $V$ can be written as a contact interaction~\cite{AlFiky:2005jd, HidalgoDuque:2012pq, Guo:2013sya, Albaladejo:2015dsa}
\begin{equation}
V(s) = A^{-1} V_d(s) A~,
\end{equation}
where $V_d(s) = \mathrm{diag}(\tC_{0Z},\tC_{0X},\tC_{1Z},\tC_{1X})$ is a diagonal matrix (the notation for the matrix elements will be explained below),\footnote{Note that the $\tC$ low-energy constants here are dimensionless, while those introduced in \cite{AlFiky:2005jd, HidalgoDuque:2012pq, Guo:2013sya, Albaladejo:2015dsa} have dimensions of fm$^2$. This is because here we adopt relativistic $D^{(*)}-$meson propagators and  non-relativistic kinematics was used in the previous works.} and the matrix $A$ (satisfying $A^T = A^{-1} = A$) reads:
\begin{equation}
A = \frac{1}{2}\left(\begin{array}{cccc}
+1 & +1 & +1 & +1 \\
+1 & -1 & +1 & -1 \\
+1 & +1 & -1 & -1 \\
+1 & -1 & -1 & +1
\end{array} \right)~.
\end{equation}
The $\mathcal{G}(s)$ matrix is diagonal, and contains the loop-function for the different two-meson channels in the particle basis, Eq.~\eqref{eq:particlebasis},
\begin{equation}\label{eq:gloopi}
\mathcal{G}_i(s) = i\int\!\! \frac{d^4 q}{(2\pi)^4} D_{Y_i}(P-q) D_{Y'_i}(q)~.
\end{equation}
where $D_{Y_i}$ and $D_{Y'_i}$ are the propagators of the two mesons $Y_i$ and $Y'_i$ in the particle basis, Eq.~\eqref{eq:particlebasis}, and $P^2=s$. In terms of the  self-energies $\Pi_Y(q)$ of the latter, they can be written as:
\begin{align}
 D_Y(q) & = \frac{1}{(q^0)^2-\omega_Y^2(\vec{q}^{\,2}) - \Pi_Y(q^0,\vec{q}\,)}  = \int_0^\infty d\omega \left( 
 \frac{S_Y(\omega,\lvert \vec{q}\, \rvert)}{q^0 - \omega + i\varepsilon} - 
 \frac{S_{\bar{Y}}(\omega,\lvert \vec{q}\, \rvert)}{q^0 + \omega - i\varepsilon}
 \right)~ 
\end{align}
with $\omega_Y(\vec{q}^{\,2})= \sqrt{m_Y^2+\vec{q}^{\,2}}$. Note that we will be here only interested in the nuclear medium renormalization of the meson properties. Thus,  $m_Y$ is the meson mass in the free space, while the self-energy $\Pi_Y$ approaches zero when the nuclear density $\rho \to 0$. Inserting the above representation into Eq.~\eqref{eq:gloopi} and integrating over $q^0$ leads to:
\begin{subequations}\label{eqs:Gifuns}\begin{equation}
\mathcal{G}_i(P^0,\vec{P}\,) = \frac{1}{2\pi^2}\int_0^\infty  \mathrm{d}\Omega \left( \frac{f_{Y_i\ Y'_i}(\Omega,\lvert \vec{P} \rvert)}{P^0 - \Omega + i\varepsilon} - \frac{f_{\overline{Y}_i\ \overline{Y}'_i}(\Omega,\lvert \vec{P} \rvert)}{P^0 + \Omega- i\varepsilon} \right)~,
\end{equation}
with:
\begin{equation}\label{eq:fUW}
f_{UW}(\Omega,\lvert \vec{P}\, \rvert) = \frac{1}{4\pi} \int_0^\Lambda  \mathrm{d}^3\vec{q}\,\int_0^\Omega \mathrm{d}\omega S_U(\omega,\lvert \vec{P}-\vec{q}\, \rvert) S_W(\Omega-\omega,\lvert \vec{q}\, \rvert\,)~,
\end{equation}\end{subequations}
where $U$ and $W$ stand for $Y_i$ and $Y_{i}'$ or $\overline{Y}_i$ and $\overline{Y}_i'$. In the previous equations we have already introduced a sharp momentum cut-off $\Lambda$ to regularize the UV behavior of the integration over the modulus of $\vec{q}$. Specifically, we take $\Lambda = 0.7\ \text{GeV}$. 

\subsection{Vacuum}
In vacuum, assuming isospin symmetry, $m_{D^{(*)0}} = m_{D^{(* +)}}$, the loop functions for the four channels are equal, $\mathcal{G}(s) = \Sigma_0(s) \mathbb{I}_{4}$. The function $\Sigma_0(s)$ reduces to a standard loop function regulated via a hard cutoff $\Lambda$, $G(s,m_D,m_{D^*})$, and expressions for this can be found in Ref.~\cite{Oller:1998hw}.
The $T$-matrix diagonalizes in the same way as the kernel matrix $V(s)$, {\it i.e.},
\begin{equation}
T^{-1}(s) = A^{-1} T_d^{-1}(s) A~,
\end{equation}
where:
\begin{equation}
T_d^{-1}(s) = \mathrm{diag}\left(\tC_{0Z}^{-1}-\Sigma_0(s),\tC_{0X}^{-1}-\Sigma_0(s),\tC_{1Z}^{-1}-\Sigma_0(s),\tC_{1X}^{-1}-\Sigma_0(s)\right)~.
\end{equation}
From the eigenvectors of $T$ (and $V$) one notices that they are $D\bar{D}^*$ states with well defined isospin $I$ ($I=0,1$) and ${\cal C}$-parity (charge-conjugation) quantum numbers. The notation $\tC_{I{\cal C}}$ for the low energy constants refers to the potential in each of these channels, with isospin and the charge-conjugation ${\cal C}=+(-)$  associated with the subindex $X$($Z$).

We consider the $X(3872)$ as a $J^P = 1^+$, $I^{\cal C} = 0^+$ state, which is thus associated to the amplitude $T_{0X}$,
\begin{equation}
T_{0X}^{-1}(s) = \tC_{0X}^{-1} - \Sigma_0(s)~.
\end{equation}
We can thus fix the constant $\tC_{0X}$ by requiring the presence of a pole at an energy equal to the $X(3872)$ mass $m_0$,\footnote{Note that, due to the regularization procedure, we should actually write $C\tC_{0X}(\Lambda) = \Sigma_0(m_0^2; \Lambda)$. For the sake of brevity, we omit this  dependence on the UV cutoff throughout the manuscript.}
\begin{equation}\label{eq:C0Xfixing}
\tC_{0X} = 1/\Sigma_0(m_0^2)~.
\end{equation}
We will also consider below (see Subsec.~\ref{subsec:tmatextension}) more general scenarios, in which energy dependence will be actually allowed in the kernel $V(s)$.

\subsection{Isospin--symmetric nuclear matter}

To consider the possible modification of the $X(3872)$ properties in a nuclear medium, we assume that the $D\bar{D}^*$ interaction potentials $V_d(s)$ do not change in nuclear matter,\footnote{This approximation is justified because they are short range (contact) interactions.} and that the $T$-matrix is modified through the loop functions because of the $D^{(*)}$ and $\bar{D}^{(*)}$ self-energies. We still assume isospin symmetry, $m_{D^{(*)0}} = m_{D^{(*)+}}$, and $S_{D^{(*)+}} = S_{D^{(*)0}} \equiv S_{D^{(*)}}$, $S_{D^{(*)-}} = S_{\bar{D}^{(*)0}} \equiv S_{\bar{D}^{(*)}}$. However, in general we will have $S_{\bar{D}^{(*)}} \neq S_{D^{(*)}}$ in the nuclear environment, since the charmed and anti-charmed meson--nucleon interactions are quite different.\footnote{Note for example that a $\bD N$ resonance would imply a pentaquark-like structure.} In addition, pseudo-scalar-nucleon and vector-nucleon interactions are also different and hence $S_D\neq S_{D^*}$ and  $S_{\bar D}\neq S_{\bar D^*}$. We discuss the spectral functions $S_{\bar{D}^{(*)}}$ and $S_{D^{(*)}}$ in nuclear matter  in Sec.~\ref{sec:DNmedium}. Consequently, the $\mathcal{G}$-matrix in a nuclear medium of density $\rho$, $\mathcal{G}(s; \rho)$, is no longer proportional to the identity, as opposed to the vacuum case. It reads $\mathcal{G}(s; \rho) = \mathrm{diag} \left(
\mathcal{G}_{D\bar{D}^*}(s; \rho),
\mathcal{G}_{D^*\bar{D}}(s; \rho),
\mathcal{G}_{D\bar{D}^*}(s; \rho),
\mathcal{G}_{D^*\bar{D}}(s; \rho)\right)$. Hence, the in-medium $T$-matrix $T(s; \rho)$ cannot be fully diagonalized, and it can only be put in block diagonal form,
\begin{equation}
T^{-1}(s;\rho ) = V^{-1}(s) - \mathcal{G}(s; \rho) = A \left( V_d^{-1}(s) - A \mathcal{G}(s; \rho) A \right) A~,
\end{equation}
with
\begin{equation}
A \mathcal{G}(s; \rho) A = \left(\begin{array}{cc} \widetilde{\mathcal{G}}(s; \rho) & 0 \\
0 & \widetilde{\mathcal{G}}(s; \rho)
\end{array}\right)~.
\end{equation}
The $2 \times 2$ matrix $\widetilde{\mathcal{G}}$ can be written as:
\begin{equation}
\widetilde{\mathcal{G}}(s; \rho) = \left(\begin{array}{cc}
\Sigma(s; \rho) & \delta_\mathcal{G}(s; \rho) \\
\delta_\mathcal{G}(s; \rho) & \Sigma(s; \rho) 
\end{array}\right)~,
\end{equation}
with:
\begin{equation}
\Sigma(s; \rho) = \frac{\mathcal{G}_{D\bar{D}^*}(s; \rho) + \mathcal{G}_{D^*\bar{D}}(s; \rho)}{2}~, \label{eq:defSigma}
\end{equation}
and:
\begin{equation}\delta_\mathcal{G}(s; \rho) = \frac{\mathcal{G}_{D\bar{D}^*}(s; \rho) - \mathcal{G}_{D^*\bar{D}}(s; \rho)}{2}~.
\end{equation}
In other words, defining the states $\ket{I {\cal C}}$, we have the following matrix elements:
\begin{equation}\label{eq:TICIpCp}
\braketO{I' {\cal C}'}{\hat{T}(s;\rho)}{I {\cal C}} = \delta_{I,I'} T^{(I)}_{{\cal C},{\cal C}'}(s; \rho)
\end{equation}
The amplitudes $T^{(I)}_{{\cal C},{\cal C}'}$ are compactly defined as:
\begin{subequations}\label{eq:Tcparity}\begin{align}
\left[T^{(I)}_{XX}(s; \rho)\right]^{-1} & = \left[T^{(I)}_X(s;\rho)\right]^{-1} - T^{(I)}_Z(s; \rho) \delta^2_\mathcal{G}(s; \rho)~,\label{eq:TIZ}\\
\left[T^{(I)}_{ZZ}(s; \rho)\right]^{-1} & = \left[T^{(I)}_Z(s;\rho)\right]^{-1} - T^{(I)}_X(s;\rho) \delta^2_\mathcal{G}(s;\rho)~, \label{eq:TIX}\\
\left[T^{(I)}_{XZ}(s; \rho)\right]^{-1} & = \left[\delta_\mathcal{G}(s;\rho) T^{(I)}_X(s; \rho)T^{(I)}_Z(s; \rho)\right]^{-1} - \delta_\mathcal{G}(s;\rho)  ~, \label{eq:TIXZ}
\end{align}\end{subequations}
where $T_X(s; \rho)$ and $T_Z(s; \rho)$ are written as in the diagonal case,
\begin{subequations}\label{eqs:medT}\begin{align}
\left[T^{(I)}_X(s;\rho)\right]^{-1} = \tC_{I X}^{-1} - \Sigma(s;\rho)~,\\
\left[T^{(I)}_Z(s;\rho)\right]^{-1} = \tC_{I Z}^{-1} - \Sigma(s; \rho)~.
\end{align}\end{subequations}
Equation~\eqref{eq:TICIpCp} may seem counter intuitive, due to the absence of a $\delta_{\cal C, \cal C'}$ factor. However, we must bear in mind that, in the presence of nuclear matter, the scattering processes are $D\bar{D}^* N \to D\bar{D}^* N'$. Due to the presence of the nucleons, the $D\bar{D}^*$ in the initial and final states do not need to have the same ${\cal C}$-parity.

We have checked that the term $\delta_\mathcal{G}(s;\rho)$ is small, so we consider throughout this manuscript the limit $\delta_\mathcal{G}(s;\rho) \to 0$. In this limit, $T_{XZ}^{(I)}(s;\rho)=0$ [Eq.~\eqref{eq:TIXZ}], and Eq.~\eqref{eq:TICIpCp} is further diagonalized into $D\bar D^*$ ${\cal C}$-parity amplitudes, too. We thus find, for the $I^{\cal C} = 0^+$ channel,\footnote{Since from now on-wards the focus will be exclusively on this channel, we will omit the $0X$ subindex for simplicity.}
\begin{equation}\label{eq:Trho_basic}
T^{-1}(s; \rho) = \tC_{0X}^{-1} - \Sigma(s; \rho)~.
\end{equation}
Note that, from its definition, $\Sigma(s; \rho)$ can be written more compactly as:
\begin{eqnarray}
\Sigma(P^0,\lvert \vec{P} \rvert; \rho) &=& \frac{1}{4\pi^2} \int_0^\infty \mathrm{d}\Omega \left(\frac{1}{P^0 - \Omega + i\varepsilon} - \frac{1}{P^0 + \Omega - i\varepsilon} \right)\nonumber \\
&\times &\left( f_{D\bar{D}^*}(\Omega,\lvert \vec{P} \,\rvert) + f_{D^* \bar{D}}(\Omega,\lvert \vec{P}\, \rvert) \right) \label{eq:Sigmarho_basic}
\end{eqnarray}
where the dependence on the density arises from that of the spectral functions involved in the above equation. We recall that the expressions for $f_{D\bar{D}^*}$, $f_{D^*\bar{D}}$ are given in Eq.~\eqref{eq:fUW}. Finally, we note that in the $\rho \to 0$ limit, the vacuum amplitudes are recovered.

In principle, given the integral representation in Eq.~\eqref{eq:Sigmarho_basic}, the function $\Sigma(P^0,\lvert \vec{P} \rvert; \rho)$ could be computed for complex values of the energy $P^0$. However, we can neither perform its analytical continuation into the lower half of the complex plane, nor define the second Riemann sheet for finite densities. This is because it would require to know the meson spectral functions $S_{U,W}$ for complex values of its arguments, which cannot be computed within the standard scheme that will be presented below, see Subsec.~\ref{sec:DNmedium}. Nevertheless, as discussed below in Subsec.~\ref{subsec:poles}, we will derive a reasonable approximation for the in-medium loop-function $\Sigma(P^0,\lvert \vec{P} \rvert; \rho)$ of Eq.~\eqref{eq:defSigma}, which will allow for a meaningful extension of the isoscalar $T$-matrix to the complex plane and the search for poles also in nuclear matter.  

\subsection[$S_{D^{(*)}}$ and $S_{\bar{D}^{(*)}}$ in nuclear matter]{\boldmath $S_{D^{(*)}}$ and $S_{\bar{D}^{(*)}}$ in nuclear matter}
\label{sec:DNmedium}

The spectral functions of $D^{(*)}$ and $\bar{D}^{(*)}$
in symmetric nuclear matter are obtained following a unitarized self-consistent procedure in coupled channels, as described in
Refs.~\cite{Tolos:2009nn,GarciaRecio:2010vt} for the $D^{(*)}$ meson and in
Ref.~\cite{GarciaRecio:2011xt} for $\bar{D}^{(*)}$ meson. In the following we present the main features.

The $s$-wave transition charmed meson--nucleon kernel of the
Bethe-Salpeter equation (BSE) is derived from an effective Lagrangian that
implements HQSS \cite{Isgur:1989vq,Neubert:1993mb,Manohar:2000dt}.  HQSS is an approximate QCD
symmetry that treats on equal footing heavy pseudoscalar and vector
mesons, such as charmed and bottomed mesons
\cite{GarciaRecio:2008dp,Gamermann:2010zz,Romanets:2012hm,GarciaRecio:2012db,Garcia-Recio:2013gaa,Tolos:2013gta,Tolos:2009nn,GarciaRecio:2010vt,GarciaRecio:2011xt,Xiao:2013yca,Ozpineci:2013qza}. The
effective Lagrangian accounts for the lowest-lying pseudoscalar and vector
mesons as well as $1/2^+$ and $3/2^+$ baryons. It reduces to the Weinberg-Tomozawa (WT)
interaction term in the sector where Goldstone bosons are involved and
incorporates HQSS in the sector where heavy quarks participate. Thus, it is a 
SU(6)$\times$HQSS model, that is justified in view of the reasonable
semi-qualitative outcome of the SU(6) extension
\cite{Gamermann:2011mq} and on a formal plausibleness on how the SU(4)
WT interaction in the heavy pseudoscalar meson-baryon sectors comes
out in the vector-meson exchange picture (see for instance Refs.~\cite{Lutz:2005ip, Mizutani:2006vq}).

This extended WT meson-baryon potential in the coupled meson-baryon
basis with total charm $C$, strangeness $S$,
isospin $I$ and spin $J$, is given by
\begin{equation}\label{eq:pot}
v_{ij}^{CSIJ}(\sqrt{t}) = D_{ij}^{CSIJ}\,\frac{2\sqrt{t}-M_i-M_j}{4f_if_j} 
\sqrt{\frac{E_i+M_i}{2M_i}}\sqrt{\frac{E_j+M_j}{2M_j}}~, 
\end{equation}
where $\sqrt{t}$ is the center of mass (C.M.) energy of the meson-baryon system; $E_i$ and
$M_i$ are, respectively, the C.M. on-shell energy and mass of the baryon in the channel
$i$; and $f_i$ is the decay constant of the meson in the $i$-channel.
Symmetry breaking effects are introduced by using physical masses and decay
constants.  The $D_{ij}^{CSIJ}$ are the matrix elements coming from the group structure of
the extended WT interaction.  

The amplitudes in nuclear matter, $t^{\rho,CSIJ}(R^0,\vec{R}\,)$ with $R=(R^0, \vec{R}\,)$ the total 
meson-baryon  four-momentum ($t=R^2)$, are obtained by solving the on-shell BSE using the previously described potential,
$v^{CSIJ}(\sqrt{t})$: 
\begin{equation}\label{eq:scat-rho}
t^{\rho,CSIJ}(R) = \left[ 1-v^{CSIJ}(\sqrt{t})\,  g^\rho_{CSIJ}(R)\right]^{-1} v^{CSIJ}(\sqrt{t})~,
\end{equation}
where the diagonal $g^\rho_{CSIJ}(R) $ loop-matrix accounts for the
charmed meson--baryon loop in nuclear matter \cite{Tolos:2009nn,GarciaRecio:2011xt}.
We focus in the non-strange $S=0$ and singly charmed
$C=1$ sector, where $DN$ and $D^*N$ are embedded, as well as the $C=-1$ one, with $\bD N$ and $\bD^* N$.\footnote{Note that $D$ denotes $D^+$ and $D^0$, whereas $\bD$ indicates $D^-$ and $\bD^0$.}

The $D(\bD)$ and $D^*(\bD^*)$ self-energies in symmetric nuclear matter, $\Pi(E,\vec{q}\,; \rho)$, are  obtained by summing
the different isospin transition amplitudes for $D(\bD)N$ and $D^*(\bD^*)N$ over the nucleon Fermi
distribution, $p_F$. For the $D(\bD)$  we have
\begin{equation}\label{eq:selfd}
\Pi_{D(\bD)}(q^0,\vec{q};\, \rho) = \int_{p \leqslant p_F} 
\frac{d^3p}{(2\pi)^3} \,
   \Big[\, t^{\rho,0,1/2}_{D(\bD) N} (R^0,\vec{R}\,) +
3 \, t^{\rho,1,1/2}_{D(\bD) N}(R^0,\vec{R}\,) \Big]~,
\end{equation}
while for $D^*(\bD^*)$
\begin{align}\label{eq:selfds}
\Pi_{D^*(\bD^*)}(q^0,\vec{q}\,; \rho\,) = \int _{p \leqslant p_F} \frac{d^3p}{(2\pi)^3} \,
& \Bigg[ \frac{1}{3} \, t^{\rho,0,1/2}_{D^*(\bD^*) N}(R^0,\vec{R}\,) +
t^{\rho,1,1/2}_{D^*(\bD^*) N}(R^0,\vec{R}\,) 
+ \\
&  \frac{2}{3} \,
t^{\rho,0,3/2}_{D^*(\bD^*) N}(R^0,\vec{R}\,) + 
2 \, t^{\rho,1,3/2}_{D^*(\bD^*) N}(R^0,\vec{R}\,)\Bigg]~. \nonumber
\end{align}
In the above equations, $R^0=q^0+E_N(\vec{p}\,)$ and $\vec{R}=(\vec{q}+\vec{p}\,)$ are the total energy and momentum of the
meson-nucleon pair in the nuclear matter rest frame, and $(q^0,\vec{q}\,)$ and $(E_N,\vec{p}\,)$ stand for the energy and momentum of the
meson and nucleon, respectively, in that frame. Those self-energies are determined self-consistently since they
are obtained from the in-medium amplitudes which contain the meson-baryon loop
functions, and those quantities themselves are functions of the self-energies.

The $D(\bD)$ and $D^*(\bD^*)$ spectral functions are then
defined from the in-medium $D(\bD)$ and $D^*(\bD^*)$  meson propagators:
\begin{eqnarray}
D^\rho_{D(\bD), D^*(\bD^*)}(q^0,\vec{q}\,)
&=&
\left ((q^0)^2 -\vec{q\,}^2-m^2-\Pi_{D(\bD),D^*(\bD^*)}(q) \right )^{-1}
,
\nonumber \\
S_{D(\bD), D^*(\bD^*)}(q^0,\vec{q}) &=& -\frac{1}{\pi}\,{\rm Im} D^\rho_{D(\bD), D^*(\bD^*)}(q)
\quad \mbox{(for~$q^0>0$)}
.
\label{eq:Drho}
\end{eqnarray}

\begin{figure*}\begin{center}
\begin{tabular}{cc}
\includegraphics{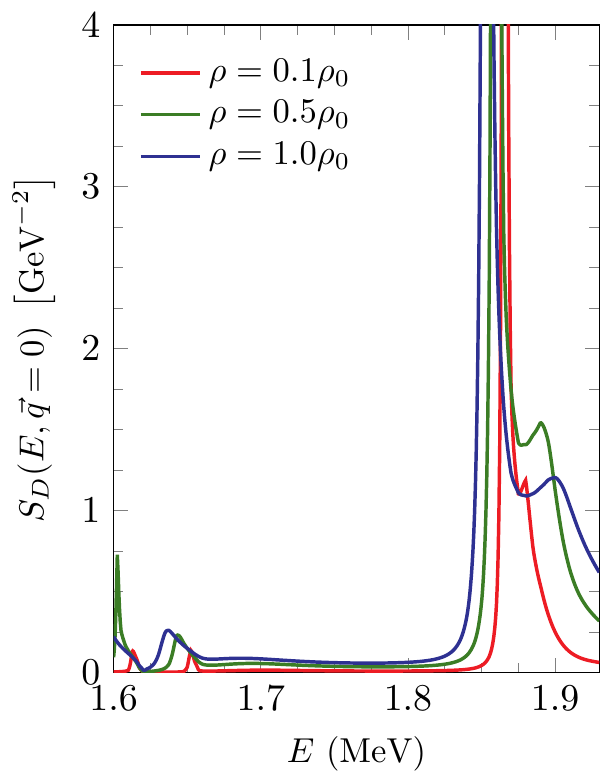} & 
\includegraphics{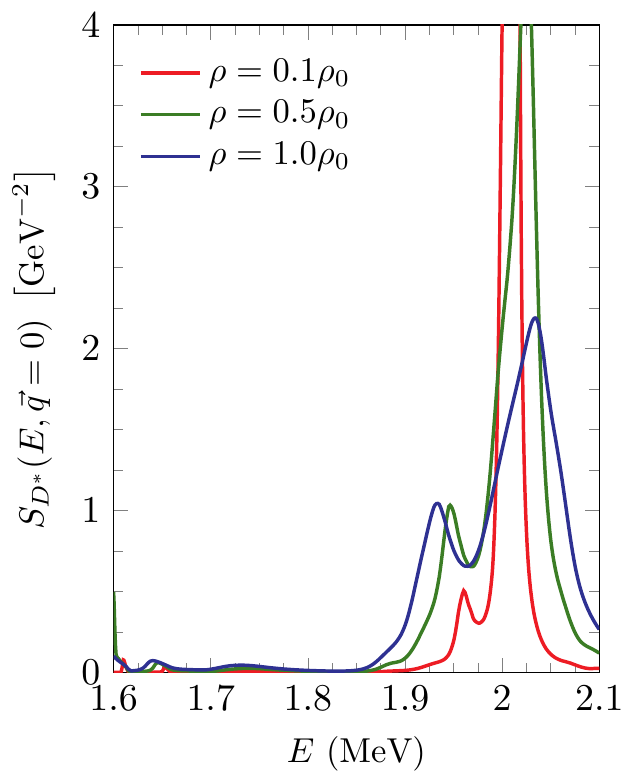} \\
\includegraphics{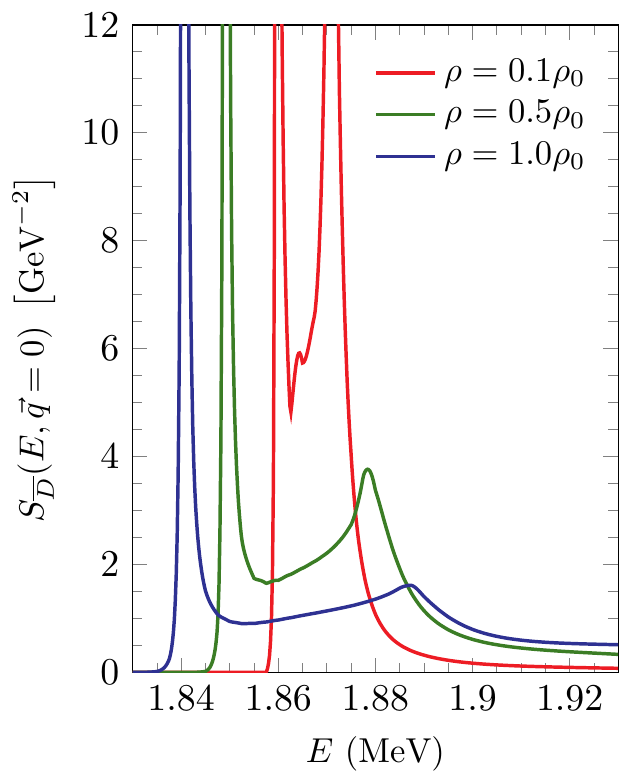} & 
\includegraphics{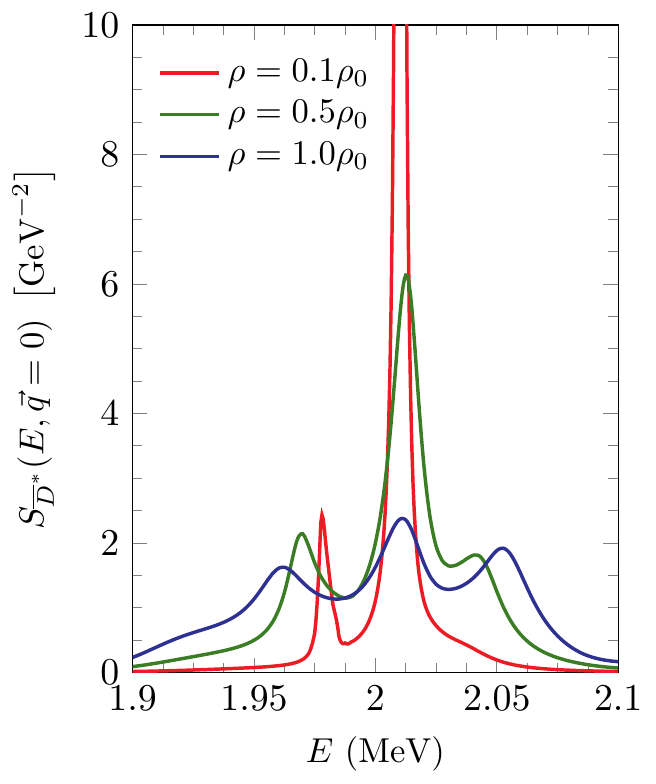}
\end{tabular}\end{center}
\caption{The $D$ (upper left-hand side), $\bD$ (lower left-hand side), $D^*$ (upper right-hand side) and $\bD^*$ (lower right-hand side) spectral functions as function of the meson energy $E$ and zero momentum $\vec{q}=0$ for two densities $\rho=0.5\rho_0$ (green lines) and $\rho=\rho_0$ (blue lines). \label{fig:SEDmesons}}
\end{figure*}

The $D(\bar D)$ and $D^*(\bar D^*)$ spectral functions are shown in Fig.~\ref{fig:SEDmesons} as function of the meson energy $E=q^0$ for zero momentum $\vec{q}=0$ and two different densities, $\rho=0.5\rho_0$ and $\rho=\rho_0$. Apart from the quasiparticle peak, obtained from  $E_\text{qp}^2=\vec{q}\,^2+m^2+{\rm Re}\Pi(E_\text{qp}(\vec{q}\,),\vec{q}\,)$, with $m$ the meson mass, these spectral functions show a rich structure as a result of the presence of several resonance-hole excitations. The masses and widths of these resonances were obtained in Refs.~\cite{GarciaRecio:2008dp,Gamermann:2010zz,Romanets:2012hm}.

The $D$ meson spectral function is depicted in the upper left-hand side panel. As described in Ref.~\cite{Tolos:2009nn}, the $D$ meson quasiparticle peak moves to lower energies with respect to the free mass position as density increases. Moreover, several resonant-hole excitations appear around the quasiparticle peak. In the low-energy tail of the $D$ spectral function, we observe the $\Lambda_c(2556)N^{-1}$ and $\Lambda_c(2595)N^{-1}$ excitations, whereas $\Sigma_c^* N^{-1}$ excitations  appear on the right-hand side of the quasiparticle peak. 

With regards to the $D^*$ meson spectral function shown in Ref.~\cite{Tolos:2009nn} and depicted here in the right-hand side panel, the quasiparticle peak moves to higher energies with density and fully mixes with the sub-threshold $J=3/2$ $\Lambda_c(2941)$ state, while the mixing of $J=1/2$ $\Sigma_c(2868)N^{-1}$ and $J=3/2$ $\Sigma_c(2902) N^{-1}$ is seen on the left-hand side of the peak. Other dynamically-generated particle-hole states appear for higher and lower energies.

Finally, the $\bD$ and $\bD^*$ spectral functions are shown in the lower left-hand side panel and lower right-hand side one, respectively.  In both cases, the spectral functions show a rich structure due to the presence of several resonance-hole states. Note that those resonant states have a pentaquark-like content and have to be taken with caution.

On the one hand, the spectral function for $\bD$ stems from the self-energy of $\bD$ displayed in Ref.~\cite{GarciaRecio:2011xt}. The position of the quasiparticle peak of $\bD$ is located below the $\bD$ mass and below the $\Theta_c(2805) N^{-1}$ excitation. The $C=-1$ pentaquark-like resonance $\Theta_c(2805)$  was a theoretical prediction of Ref.~\cite{Gamermann:2010zz}. This corresponds to a pole in the free space amplitude of the sector $I=0$,$J=1/2$ (a weakly bound state) that strongly couples to $\bar D N$ and $\bar D^* N$, also found in Ref.~\cite{Yasui:2009bz}, though it has not been observed yet. 

The upper energy tail of the $\bD$ spectral function shows also the contribution of $I=1$ resonant-hole states. On the other hand, the $\bD^*$ spectral function depicts the contribution of several $I=0$ and $I=1$ resonant-hole states close to the quasiparticle peak, that is located slightly above to 2 GeV. All these pentaquark-like states are described in Ref.~\cite{Gamermann:2010zz}.

\section{\boldmath Self-energy formalism and extension of the $T$-matrix formalism.}
\label{sec:self-energy}

From this section on, and since we focus on the $I^{\cal C} = 0^+$ channel, where the $X(3872)$ is located, for $D \bar D^* $ we mean the appropriate combination of states, $(D \bar D^*-D^* \bar D)\sqrt{2}$, with even ${\cal C}-$parity and coupled to zero isospin.

\subsection[$X(3872)$ self-energy in vacuum and in a nuclear medium]{\boldmath $X(3872)$ self-energy in vacuum and in a nuclear medium}
\label{subsec:only_self_energy}
\begin{figure}\centering
\begin{tabular}{cc}
\includegraphics{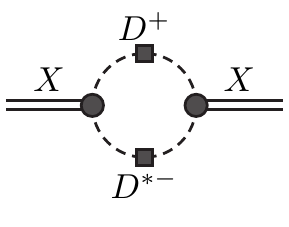} & 
\includegraphics{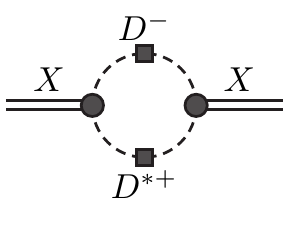} \\
\includegraphics{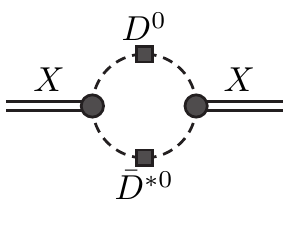} & 
\includegraphics{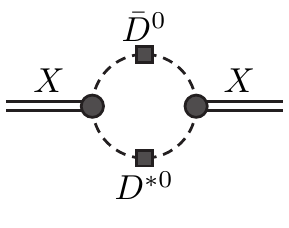}
\end{tabular}
\caption{Contributions to the $X(3872)$ self-energy in nuclear matter. Circles represent the $X(3872)$ couplings to the meson pairs, and the squares the interaction of the charm mesons with nuclear matter.\label{fig:diags}}
\end{figure}
We shall now discuss the self-energy formalism for the $X(3872)$. Let us consider a ``pre-existing'' state with bare mass $\hat{m}$ and bare coupling squared to each of the four channels $\hat{g}^2/4$. (The isospin related factor $1/4$ is included for convenience.) The free-space bare propagator $\hat{\Delta}(q^2)$ is:
\begin{equation}
\hat{\Delta}^{-1}(q^2) = q^2 - \hat{m}^2 + i\varepsilon~.
\end{equation}
Upon resumation of the contributions in Fig.~\ref{fig:diags}, the dressed propagator reads:
\begin{equation}
\Delta^{-1}(q^2; \rho) = \hat{\Delta}^{-1}(q^2) - \hat{g}^2 \Sigma(q^2;\rho)~. 
\end{equation}
This renormalizes the mass  and  coupling of the state in the medium,
\begin{subequations}\label{eq:mgrho_a}\begin{align}
m^2(\rho) & = \hat{m}^2 + \hat{g}^2 \Sigma[m^2(\rho); \rho]~,\\
g^2(\rho) & = \frac{\hat{g}^2}{1-\hat{g}^2 \Sigma'[m^2(\rho); \rho]}~.
\end{align}\end{subequations}
with $\rho$ the nuclear-matter density as in the previous sections, and the derivative taken with respect to $q^2=s$. These equations are also true in particular for the $\rho = 0$ case, so that we can relate the bare mass and coupling to the vacuum ones, $m_0$ and $g_0$:
\begin{subequations}\label{eqs:mgbare}\begin{align}
\hat{m}^2 & = m_0^2 - \frac{g_0^2}{1+g_0^2 \Sigma'_0(m_0^2)}\Sigma_0(m_0^2)~,\\
\hat{g}^2 & = \frac{g_0^2}{1+g_0^2 \Sigma'_0(m_0^2)}~.
\end{align}\end{subequations}
This allows in turn to rewrite the in-medium mass and coupling, $m(\rho)$ and $g(\rho)$, in terms of the physical ones in vacuum:
\begin{subequations}\label{eq:rho0}\begin{align}
m^2(\rho) & = m_0^2 + \frac{g_0^2}{1+g_0^2 \Sigma'_0(m_0^2)} \left[ \Sigma[m^2(\rho); \rho] - \Sigma_0(m_0^2) \right]~, \label{eq:mrhom0}\\
g^2(\rho) & = \frac{g_0^2}{1-g_0^2 \big[ \Sigma'[m^2(\rho); \rho] - \Sigma'_0(m_0^2) \big]}~. \label{eq:grhog0}
\end{align}\end{subequations}
Note that $m^2(\rho)$ is in general a complex quantity, its imaginary part being originated by that of $\Sigma[m^2(\rho); \rho]$.\footnote{Even assuming that in the free-space the $X(3872)$ is bound, and therefore $\Sigma_0(m_0^2)$ is real, the in-medium self-energy might acquire an imaginary part since new many-body decay modes, induced by the quasi-elastic interactions of the $D^{(*)}$ and  $\bar D^{(*)}$ with nucleons,   are open. } We can also rewrite the in-medium $X(3872)$ propagator as:
\begin{align}
\Delta^{-1}(q^2; \rho) & = q^2 - m_0^2 - \frac{g_0^2}{1+g_0^2 \Sigma'_0(m_0^2)} \left( \Sigma(q^2; \rho) - \Sigma_0(m_0^2) \right) \equiv q^2 - m_0^2 - \Pi_X(q^2; \rho)~,\label{eq:proprho_basic}\\
\Pi_X(q^2; \rho) & = \frac{g_0^2}{1+g_0^2 \Sigma'_0(m_0^2)} \left( \Sigma(q^2;\rho) - \Sigma_0(m_0^2) \right)~,\label{eq:serho_basic}
\end{align}
which defines the $X(3872)$ self-energy in a nuclear medium, $\Pi_X(q^2; \rho)$. We can now rewrite Eqs.~\eqref{eq:rho0} as:
\begin{subequations}\begin{align}
m^2(\rho) & = m_0^2 + \Pi_X[m^2(\rho); \rho],\\
g^2(\rho) & = \frac{\hat{g}^2}{1-\Pi'_X[m^2(\rho); \rho]} = g_0^2 \frac{1-\Pi'_X(m_0^2;\rho=0)}{1-\Pi'_X[m^2(\rho);\rho]}~.
\end{align}\end{subequations}
Once the $X(3872)$ propagator or self-energy are known, one can also define the $X(3872)$ spectral function, $S_X(q^2;\rho)$,
\begin{equation}\label{eq:spef_basic}
S_X(q^2; \rho) = - \frac{1}{\pi}\text{Im}\Delta(q^2; \rho) = -\frac{1}{\pi}\frac{\text{Im} \Pi_X(q^2; \rho)}{\left[q^2-m_0^2 - \text{Re}\Pi_X(q^2;\rho) \right]^2 + \left[ \text{Im}\Pi_X(q^2; \rho)\right]^2}~.
\end{equation}
The quasi-particle peak energy, $E_\text{qp}$, is defined from the equation:
\begin{equation}
E_\text{qp}^2-m_0^2-\text{Re}\Pi(E^2_\text{qp}; \rho) = 0~. \label{eq:Eqp}
\end{equation}

\subsection[Extension of the $T$-matrix formalism and relation with the self-energy formalism]{\boldmath Extension of the $T$-matrix formalism and relation with the self-energy formalism}\label{subsec:tmatextension}
We now seek for a relation between the $T$-matrix and the self-energy formalism introduced in the previous subsection. We consider the in-medium $T$-matrix, $T^{-1}(s;\rho) = V^{-1}(s) - \Sigma(s; \rho)$, with a potential $V$ somewhat more general than a simple constant, $\tC_{0X}$. Specifically, we allow for a term linear in the Mandelstam variable $s$, and write:
\begin{subequations}\label{eqs:V_pot_a}
\begin{eqnarray}
\label{eq:Vlinear}
V(s) &=& \frac{1}{\Sigma_0(m_0^2)} + \frac{\Sigma'_0(m_0^2)}{\Sigma_0^2(m_0^2)} \frac{1-P_0}{P_0}(s-m_0^2)\\
&=&\frac{\hat g^2}{m_0^2-\hat m^2}-\frac{\hat g^2}{(m_0^2-\hat m^2)^2}\,\left(s-m_0^2 \right) \equiv V_A(s)~.
\end{eqnarray}
\end{subequations}
Note that $V_A(m_0^2) = 1/\Sigma_0(m_0^2)$, which is the same constant term that was previously considered, see Eq.~\eqref{eq:C0Xfixing}. Hence, with this potential, the free-space amplitude $T_0(s)$ has a pole at $s=m_0^2$,
\begin{equation}
T_0(s) \simeq \frac{g_0^2}{s-m_0^2} + \cdots~,
\end{equation}
with coupling $g_0$ given by:
\begin{equation}\label{eq:Prob_vac}
\frac{1}{g_0^2} = \left. \frac{\mathrm{d} T^{-1}_0(s)}{\mathrm{d}s} \right\rvert_{s=m_0^2} = -\frac{\Sigma'_0(m_0^2)}{P_0}~.
\end{equation}
According to the Weinberg compositeness condition~\cite{Weinberg:1965zz}, the factor $-g_0^2 \Sigma'_0(m_0^2)$ represents the $D\bar{D}^*$ component in the $X(3872)$ wave function. Hence, the linear term in the potential is chosen so as to set this probability equal to $P_0$.

If there is a pole of the amplitude $T(s; \rho)$ at $m^2(\rho)$, then:
\begin{align}
0 & = T^{-1}[m^2(\rho); \rho] = V^{-1}[m^2(\rho)] - \Sigma[m^2(\rho); \rho] \label{eq:mrhofromT} \\
& \simeq \Sigma_0(m_0^2) + \frac{1+g_0^2 \Sigma'_0(m_0^2)}{g_0^2}\left(m^2(\rho)-m_0^2\right) - \Sigma[m^2(\rho); \rho]~,\nonumber
\end{align}
from where one obtains the same equation for $m^2(\rho)$ than that obtained in Eq.~\eqref{eq:mrhom0} within the self-energy formalism. Analogously, the in medium coupling $g(\rho)$ would be given by:
\begin{align}
\frac{1}{g^2(\rho)} & = \left. \frac{\mathrm{d} T^{-1}(s; \rho)}{\mathrm{d}s} \right\rvert_{s=m^2(\rho)} \nonumber\\
& = \left[ \frac{\Sigma[m^2(\rho); \rho]}{\Sigma_0(m_0^2)} \right]^2 \frac{1 + g_0^2 \Sigma'_0(m_0^2)}{g_0^2} - \Sigma'[m^2(\rho); \rho]~.
\end{align}
This latter equation does not give exactly the same result than Eq.~\eqref{eq:grhog0} because of the factor between the square brackets. If that factor is taken as 1, one recovers Eq.~\eqref{eq:grhog0}. 

Alternatively, we could have made the linear expansion in $1/V(s)$ instead of in $V(s)$ [Eq.~\eqref{eq:Vlinear}], thus getting:
\begin{subequations}\label{eqs:V_pot_b}
\begin{equation}\label{eq:Vlinearalt}
V^{-1}(s) = \Sigma_0(m_0^2) - \Sigma'_0(m_0^2)\frac{1-P_0}{P_0}(s-m_0^2) \equiv V_B(s)~.
\end{equation}
Note that this alternate definition of $V(s)$ can also be written as:
\begin{equation}
V_B(s) = \frac{\hat g^2}{s-\hat m^2}~.
\end{equation}
\end{subequations}
{\textit i.e.}, the kernel has a ``bare'' pole at the ``bare'' mass squared $\hat m^2$.
Then we would obtain:
\begin{equation}
\frac{1}{g^2(\rho)} = \left. \frac{\mathrm{d} T^{-1}(s; \rho)}{\mathrm{d}s} \right\rvert_{s=m^2(\rho)} = \frac{1 + g_0^2 \Sigma'_0(m_0^2)}{g_0^2} - \Sigma'[m^2(\rho); \rho]~,
\end{equation}
which allows to recover Eq.~\eqref{eq:grhog0}. Equation \eqref{eq:Vlinearalt} should be a good approximation to \eqref{eq:Vlinear} for $s$ in the neighborhood of $m_0^2$ if the factor $\Sigma'_0(m_0^2)(1-P_0)/P_0$ is sufficiently small. Indeed, it has been considered also in Eq.~\eqref{eq:mrhofromT}. Hence, we find equivalence between the self-energy formalism (Subsec.~\ref{subsec:only_self_energy}) and the $T$-matrix formalism(s) presented here.

Note finally that taking into account the relation $P_0 = - g_0^2 \Sigma'_0(m_0^2)$, Eqs.~\eqref{eq:rho0} can be cast as:
\begin{subequations}\label{eq:rho0P}\begin{align}\displaystyle
m^2(\rho) & = m_0^2 - \frac{P_0}{1-P_0} \frac{\Sigma[m^2(\rho); \rho]-\Sigma_0(m_0^2)}{\Sigma'_0(m_0^2)}~,\\
g^2(\rho) & = -\frac{1}{\Sigma'[m^2(\rho); \rho] + \frac{1-P_0}{P_0}\Sigma'_0(m_0^2)}~.
\end{align}\end{subequations}

\subsection[Extreme cases: $P_0 \to 0$ and $P_0 \to 1$]{\boldmath Extreme cases: $P_0 \to 0$ and $P_0 \to 1$}

Let us briefly discuss here the extreme molecular or compact state scenarios, which correspond to $P_0 \to 1$ or $P_0 \to 0$, respectively.

We start by considering the case when $P_0 \to 0$. In this case one has $g_0 = 0$, {\it i.e.}, the state does not couple to the two-meson channel. Physically, one would say that the interaction does not renormalize the bare state. Indeed, one sees also that $\hat{g} = g_0  = g(\rho) = 0$, and that $\hat{m} = m_0 = m(\rho)$. This case is nonphysical, since it would require $V'(s=m_0) \sim 1/P_0 \to  \infty$.

Next we discuss the opposite case $P_0 \to 1$. This situation would correspond to the pure hadron-molecular case, for which $V(s) = 1/\Sigma_0(m_0^2)$ is constant, independent of $s$. The search of a pole in the nuclear-medium $T$-matrix  would lead to $\Sigma[m^2(\rho); \rho] = V^{-1}= \Sigma_0(m_0^2)$. Actually in this limiting case, 
\begin{equation}
    T(s; \rho)= \frac{1}{\Sigma_0(m_0^2)-\Sigma(s;\rho)} \label{eq:Tp0eq1}
\end{equation}
which cannot have a pole on the real axis, since  $\Sigma(s;\rho)$ is a complex magnitude, and if there exists a pole, it will be located at a complex value $\sqrt{s}=m(\rho) \in \mathbb C$ (see Subsec.~\ref{subsec:poles}), and the coupling from the residue will be given by  $g^2(\rho) = - 1/\Sigma'[m^2(\rho);\rho]$. These results would also make sense to the first of the Eqs.~\eqref{eq:rho0P}: since the denominator $1-P_0$ tends to zero, the numerator must also vanish, finding thus $ \Sigma[m^2(\rho); \rho]- \Sigma_0(m_0^2) = 0 $. Physically, taking $P_0 \to 1$ means that the state is a purely molecular one. The ``pre-existing'' component is null, and thus one can neither think about the bare mass nor about the bare coupling. Indeed, for $P_0 \to 1$, the factor $1/(1-P_0)$ diverges and so it does $1/\left( 1+g_0^2 \Sigma'_0(m_0^2) \right)$, and hence  $\hat{g} \to \infty$  and $\hat{m}\to \infty$, in Eq.~\eqref{eqs:mgbare}.

For simplicity, in the discussion above, we have not considered the pathological case in which the bound state in vacuum  is placed exactly at threshold. In that case  $\Sigma'_0(m_0^2)$ diverges, and this singular behaviour needs to be taken into account. 

\section{Results}
\label{sec:results}

\subsection{In medium modification of the amplitudes}\label{subsec:amplis}

\begin{figure}\centering
\begin{tabular}{cc}
\multicolumn{2}{c}{\includegraphics[width=0.60\textwidth]{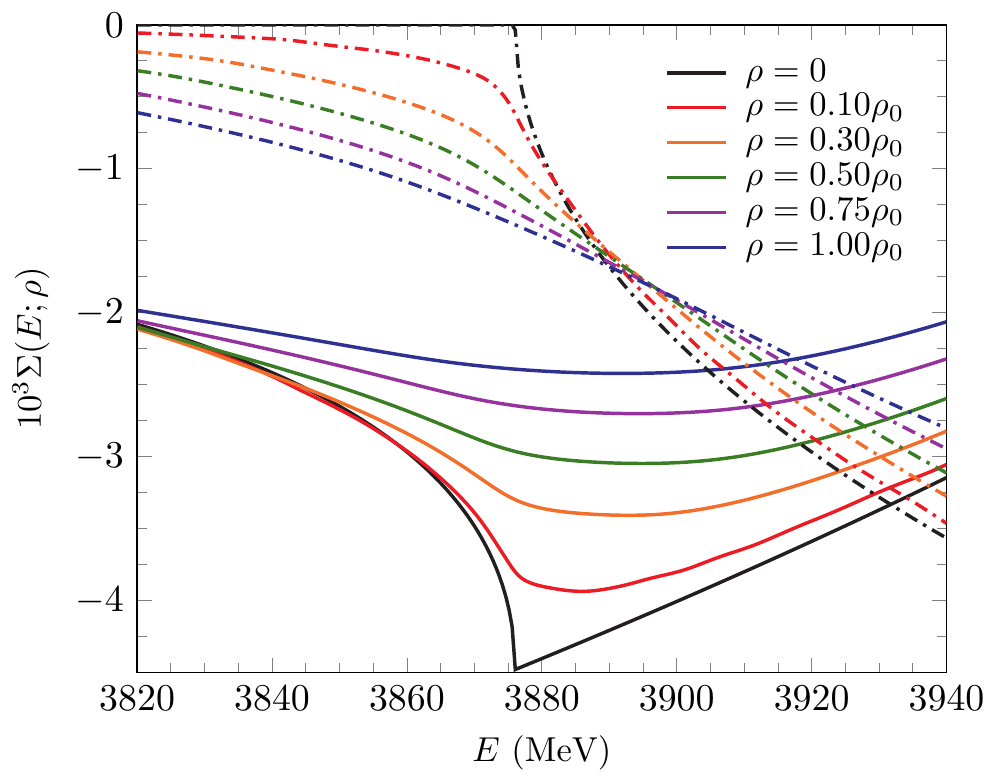}} \\
\includegraphics[width=0.40\textwidth]{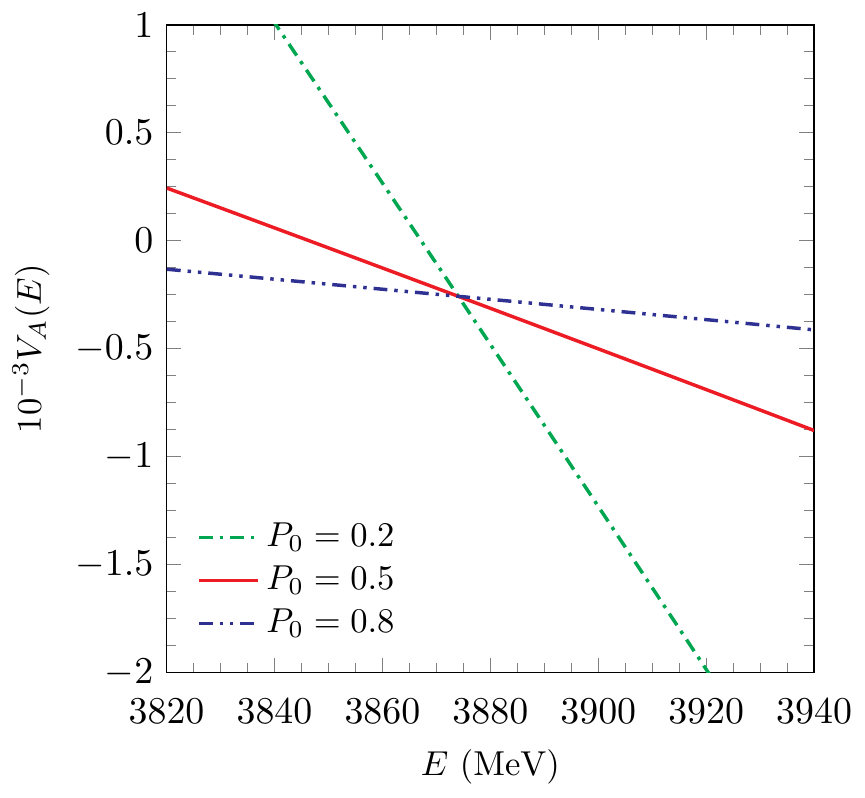} &
\includegraphics[width=0.40\textwidth]{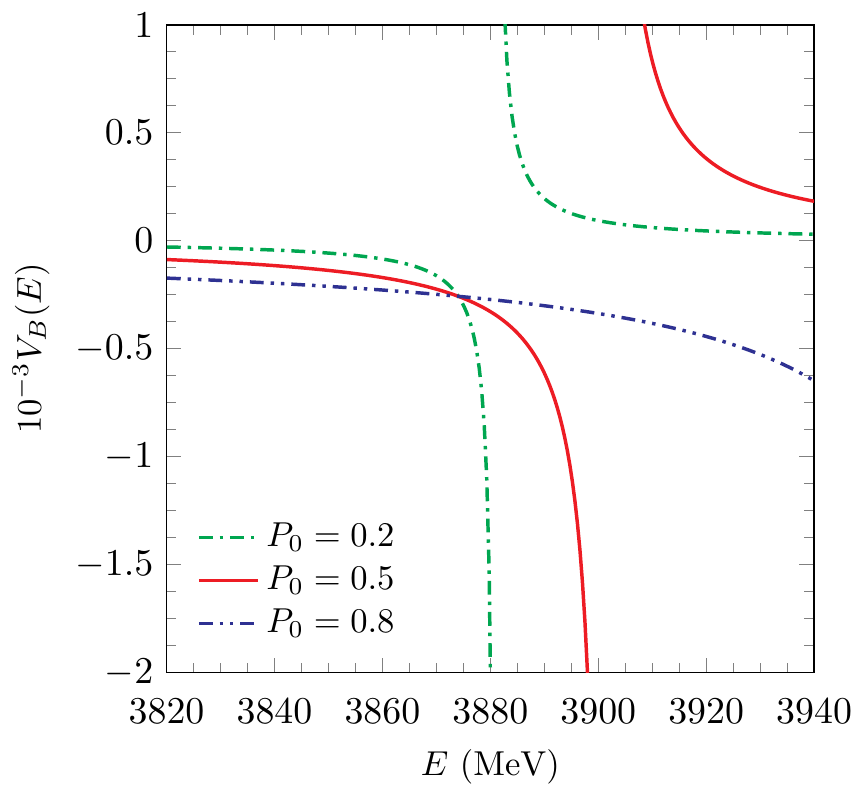}
\end{tabular}
\caption{\label{fig:sigmarho}%
Top panel: The loop function $\Sigma(E; \rho)$, with $E^2=s$, for different densities $\rho$ in the range $0 \leqslant \rho \leqslant \rho_0$ as a function of the center-of-mass energy of the $D\bar D^*$ pair. The solid (dashed) lines stand for the real (imaginary) parts. Bottom panels: Two different parameterizations of the energy-dependent potential $V(s)$. On the left [right] plot, $V_A(s)$ [$V_B(s)$], as given in Eq.~\eqref{eqs:V_pot_a} [Eq.~\eqref{eqs:V_pot_b}]~.}
\end{figure}

We now discuss the results that we obtain in a nuclear medium for the $D\bar{D}^*$ amplitude\footnote{We recall here again that we are working on the $I^{\cal C} = 0^+$ channel, where the $X(3872)$ is located, and that for $D \bar D^* $ we mean the appropriate combination of states, $(D \bar D^*-D^* \bar D)\sqrt{2}$, with even ${\cal C}-$parity and coupled to zero isospin.} $|T(E;\rho)|^2$ ({\it cf.}~Eq.~\eqref{eq:Trho_basic}, using a very general energy dependent potential instead of just a constant $\tC_{0X}$), the $X(3872)$ self-energy $\Pi_X(E; \rho)$ [cf.~Eq.~\eqref{eq:serho_basic}] (or, equivalently, the inverse of the propagator $\Delta^{-1}(E; \rho)$ [cf.~Eq.~\eqref{eq:proprho_basic}]), and its spectral function $S_X(E; \rho)$ [cf.~Eq.~\eqref{eq:spef_basic}].  Note that we  use the energy, $E$, of the $D\bar D^*$-pair in the c.m. frame, with $s=E^2$. In order to compute all these quantities, we need the energy-dependent potential $V(s)$ [{\it cf.} \eqref{eqs:V_pot_a} or \eqref{eqs:V_pot_b}] and  the in-medium modified $D\bar{D}^*$ loop function $\Sigma(s; \rho)$ [cf. Eq.~\eqref{eq:Sigmarho_basic}].
\begin{figure*}\centering
\includegraphics[width=0.95\textwidth]{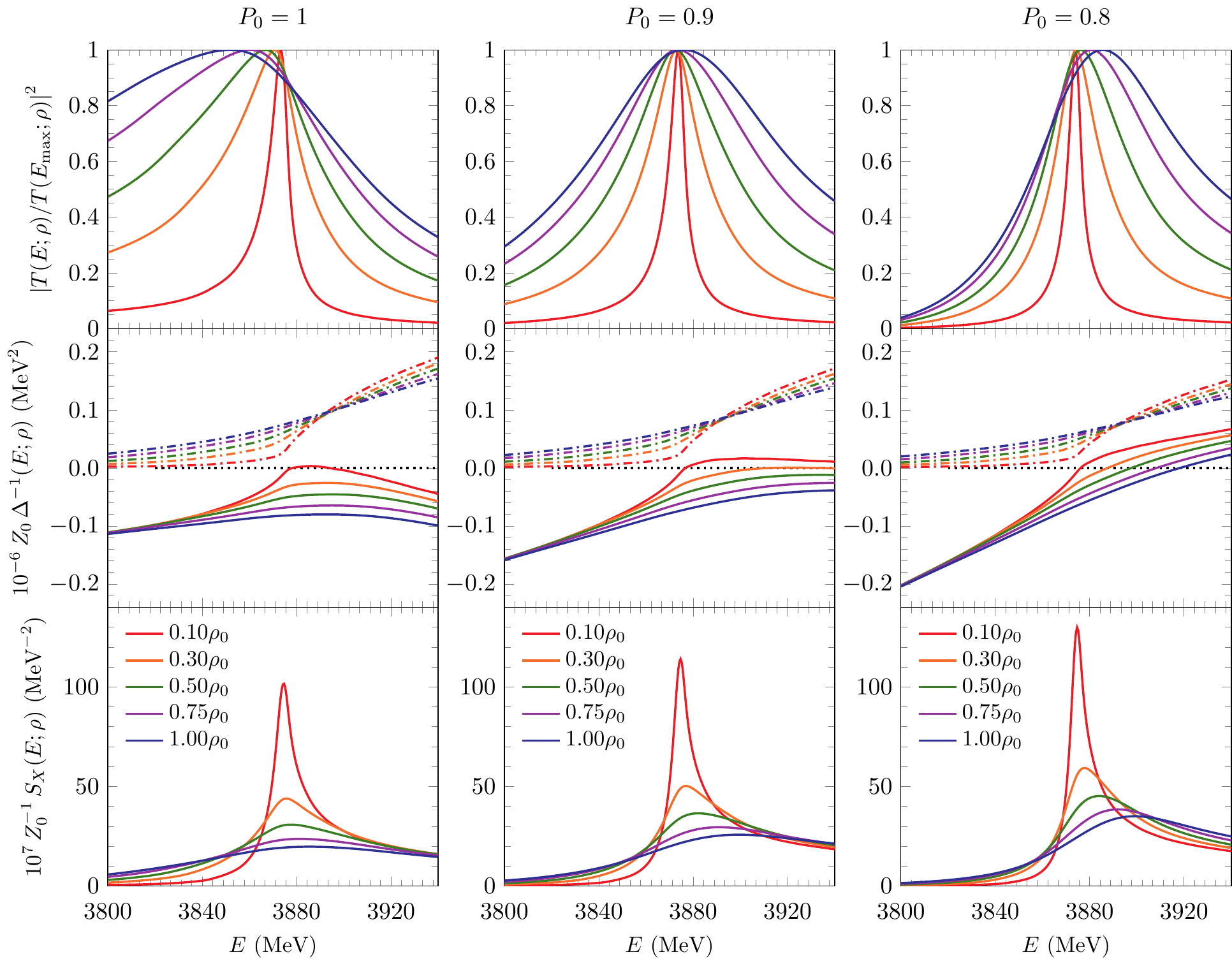}
\caption{\label{fig:ResT}Top panels: Squared modulus of the amplitudes $T(E; \rho)$, normalized to be one at the maximum, $E_\text{max}$ as a function of the energy of the $D\bar D^*$-pair in the c.m. frame. The amplitudes are computed with Eq.~\eqref{eq:Trho_basic}, using  the energy dependent potential of Eq.~\eqref{eqs:V_pot_a} instead of the constant $\tC_{0X}$. Middle panels: Real (solid lines) and imaginary parts (dashed lines) of the inverse of the propagator $\Delta(E; \rho)$ (Eq.~\eqref{eq:proprho_basic}) multiplied by $Z_0=(1-P_0)$, as a function of the energy of the $D\bar D^*$-pair in the c.m. frame. Bottom panels: Spectral function of the $X(3872)$ (Eq.~\eqref{eq:spef_basic}) multiplied by $Z_0^{-1}$, as a function of the energy of the $D\bar D^*$-pair in the c.m. frame. From left to right, the three columns show the cases $P_0=1$, $0.9$, and $0.8$. For these high molecular probabilities, the numerical differences due to the use of $V_A(s)$ or of $V_B(s)$ potentials [Eqs.~\eqref{eqs:V_pot_a} and \eqref{eqs:V_pot_b}, respectively] are very small. The different colors in each figure represent calculations performed at different nuclear densities  $0 \leqslant \rho \leqslant \rho_0$.} 
\end{figure*}
In the upper plot of Fig.~\ref{fig:sigmarho} we show the real (solid lines) and imaginary (dot-dashed lines) parts of the loop function $\Sigma(s; \rho)$ computed for different densities $\rho$ in the range $0 \leqslant \rho \leqslant \rho_0$, where $\rho_0$ is the normal nuclear density, $\rho_0 = 0.17\ \text{fm}^{-3}$. We see that the sharp $D \bar D^*$ threshold observed in the vacuum case ($\rho=0$) is progressively smoothed out for increasing densities, being almost inappreciable for $\rho = \rho_0$. This is due to the width acquired by  the $D$, $\bar D$, $D^*$ and $\bar{D}^*$ mesons in the nuclear medium. We also notice that the real part of the loop function is smaller in magnitude for increasing densities. Naively, this would imply that the effect of the medium is to generate repulsion in the $D \bar D^*$ interaction, in the sense that a more attractive potential would be necessary to compensate this change of the loop function. However, this repulsive effect is not clear, because the imaginary part of $\Sigma(E;\rho)$ is also large, and below threshold, it turns out that $ \left\lvert \text{Im} \Sigma(E; \rho) \right\rvert \gtrsim \left\lvert \text{Re} \left(\Sigma(E;\rho) - \Sigma_0(E) \right) \right\rvert$.

Within the present approach, the $D\bar{D}^*$ $T-$matrix in the nuclear environment is determined from the $X(3872)$ mass and its $D\bar{D}^*$ probability ($m_0$ and $P_0$)  in the vacuum ($\rho = 0$). As we work on the isospin limit, $m_{D^{(*)}} = \left( m_{D^{(*)+}} + m_{D^{(*)0}} \right)/2$, we cannot consider the physical $X(3872)$ mass. We instead take a binding energy $B = 2\ \text{MeV}$ with respect to the $D\bar{D}^*$ threshold, $m_0 = m_D + m_{D^*} - B$. Throughout this manuscript, we will study the in-medium effects for different molecular probabilities $P_0$, that enter into the calculation of the amplitude through the potentials  $V_A(s)$ or $V_B(s)$, Eqs.~\eqref{eqs:V_pot_a} and \eqref{eqs:V_pot_b}, respectively. Indeed, in the lower plots of Fig.~\ref{fig:sigmarho}, we show these interaction kernels computed for different values of $P_0$. Both types of interactions give the same pole position at $m_0^2$ and probability $P_0$ (alternatively, the same coupling $g_0$) for the vacuum $T$-matrix, although they have different analytical properties ($V_A(s)$ has a zero, while $V_B(s)$ presents a bare pole) and, hence, they might produce differences in the medium $T$-matrix, as we will discuss below. In the lower panels of Fig.~\ref{fig:sigmarho} we observe, on the one hand, that for values of $P_0$ above $P_0=0.8$ both kernels are very similar in the energy region explored. This is due to the fact that the zero of $V_A(s)$ and the bare pole of $V_B(s)$ are far from the energies considered. On the other hand, for lower values of $P_0$, {\it e.g.} $P_0=0.2$, both potentials are quite different, because the zero of $V_A(s)$ and the bare pole $V_B(s)$ come closer to the energy region of interest. Therefore, one should expect that they lead to significantly different in medium $T$-matrices.

Once discussed the in-medium modified $D\bar{D}^*$ loop function and the energy-dependent potential,  in Fig.~\ref{fig:ResT} we show, for different nuclear densities and molecular probabilities $P_0=1,0.9$ and 0.8, the squared modulus of the amplitudes $T(E;\rho)$, normalized to be one at the maximum $E_{\rm max}$ (top panels), the inverse of the $X(3872)$ propagator, $\Delta^{-1}(E; \rho)$ (medium panels), and the spectral function, $S_X(E; \rho)$ (bottom panels), conveniently scaled by $Z_0=(1-P_0)$ and $Z_0^{-1}$, respectively. The calculations are performed using the potential $V_A(s)$, introduced in Eqs.~\eqref{eqs:V_pot_a}, though, as shown above, for these high-molecular component scenarios the $V_B(s)-$type interaction, {\it cf.} Eqs.~\eqref{eqs:V_pot_b}, leads to very similar predictions, with differences that would be difficult to appreciate in the plots.   

 Focusing first on the squared amplitudes, it can be seen that the density behaviour is qualitatively different for the three examined probabilities. Thus, while for $P_0=0.8$ the maximum of the squared modulus is shifted to the right when the density grows (towards higher $D\bar D^*$ c.m. energies), it however moves to the left in the purely molecular ($P_0=1)$ scenarios. The results for $P_0=1$ stem from the energy and density behaviour of the factor $\left|\Sigma_0(m_0^2)-\Sigma(E; \rho)\right|$
in Eq.~\eqref{eq:Tp0eq1}, by taking into account the in-medium two-meson loop function $\Sigma(E; \rho)$ depicted in Fig.~\ref{fig:sigmarho}. For the $P_0=0.8$ case, the energy dependence of the $V_A$ potential, shown in the left-bottom plot of  Fig.~\ref{fig:sigmarho}, leads to the mild shift towards higher energies of the maximum as the density increases. The position of the peak hardly changes in the intermediate $P_0=0.9$ case, displayed in the second-column plot, but as expected, the width of the in-medium $X(3872)$ peak significantly increases with density. 

Actually, in the second row of plots of Fig.~\ref{fig:ResT}, we see that the energy dependence  of Im$\Pi_X(E;\rho)$ for finite density clearly departs from the sharp step-function shape obtained in vacuum, with Im$\Pi_X(E;\rho)$ becoming an increasingly  smoother function of $E$, as the density grows. We moreover observe non-vanishing values below the free-space threshold, which increase with the density,  due to the appearance of new many-body decay channels, like $D\bar D^* N \to D \bar D^* N'$, driven by the self-energies of the (anti)charmed mesons embedded in the nuclear medium. Above the free-space threshold, Im$\Pi_X(E;\rho)$ decreases when the density grows. This behaviour can be inferred from the imaginary part of $\Sigma(E; \rho)$ shown in the top plot of Fig.~\ref{fig:sigmarho}.

We should also note that Im$\Pi_X(E;\rho)$ strongly depends on $P_0$, and it behaves as $g_0^2/[1+g_0^2 \Sigma'_0(m_0^2)] \propto P_0/(1-P_0)$, as deduced from Eq.~\eqref{eq:serho_basic}. Looking now at real part of $\Delta^{-1}(E;\rho)$, we observe that for $P_0=1$, there is not quasi-particle solution (Eq.~\eqref{eq:Eqp}) for densities higher than about one tenth of the normal nuclear matter density, with an increasingly flatter $E-$dependence of Re$[\Delta^{-1}(E;\rho)]$ as the density grows.  Hence, the behavior exhibited in the $P_0=1$ case in left-top plot for the modulus squared of the amplitude, with the maximum  displaced to the left with increasing densities, can be correlated to  the growth of Im$\Pi_X(E;\rho)$, both with the density and the c.m. energy. On the contrary, for $P_0=0.8$, we find solutions for the quasi-particle equation for all densities, at energies above threshold that move away of it as the density increases. 

The spectral function plotted in the bottom panels  of Fig.~\ref{fig:ResT} is determined by  Im$[\Delta(E;\rho)]$, and its dependence on $E, \rho$ and the molecular probability $P_0$ can be deduced from the discussion above on the real and imaginary parts of  $\Delta^{-1}(E;\rho)$ in the second-row panels of this figure. We should make here two remarks. First, we observe that the typical delta-function shape expected for the spectral function of a narrow state in the free space gets diluted as the density grows. This is due to the enhancement of the $X(3872)$ width with density. Second, we find that, for purely molecular case $(P_0=1)$, the features of the modulus squared of the $T-$matrix (top-left plot)  can not  be inferred from the spectral function $S_X(E;\rho)$. This situation slowly changes as the molecular probability decreases. Indeed, for $P_0=0.8$, we observe already some resemblances between $|T(E;\rho)|^2$ and  $S_X(E;\rho)$. Nevertheless, the squared amplitude $|T(E;\rho)|^2$ is the observable that elucidates the properties of the $X(3872)$ in the medium, especially in cases of high (dominant) molecular components in its vacuum structure. 

\begin{figure*}\centering
\includegraphics[width=0.80\textwidth]{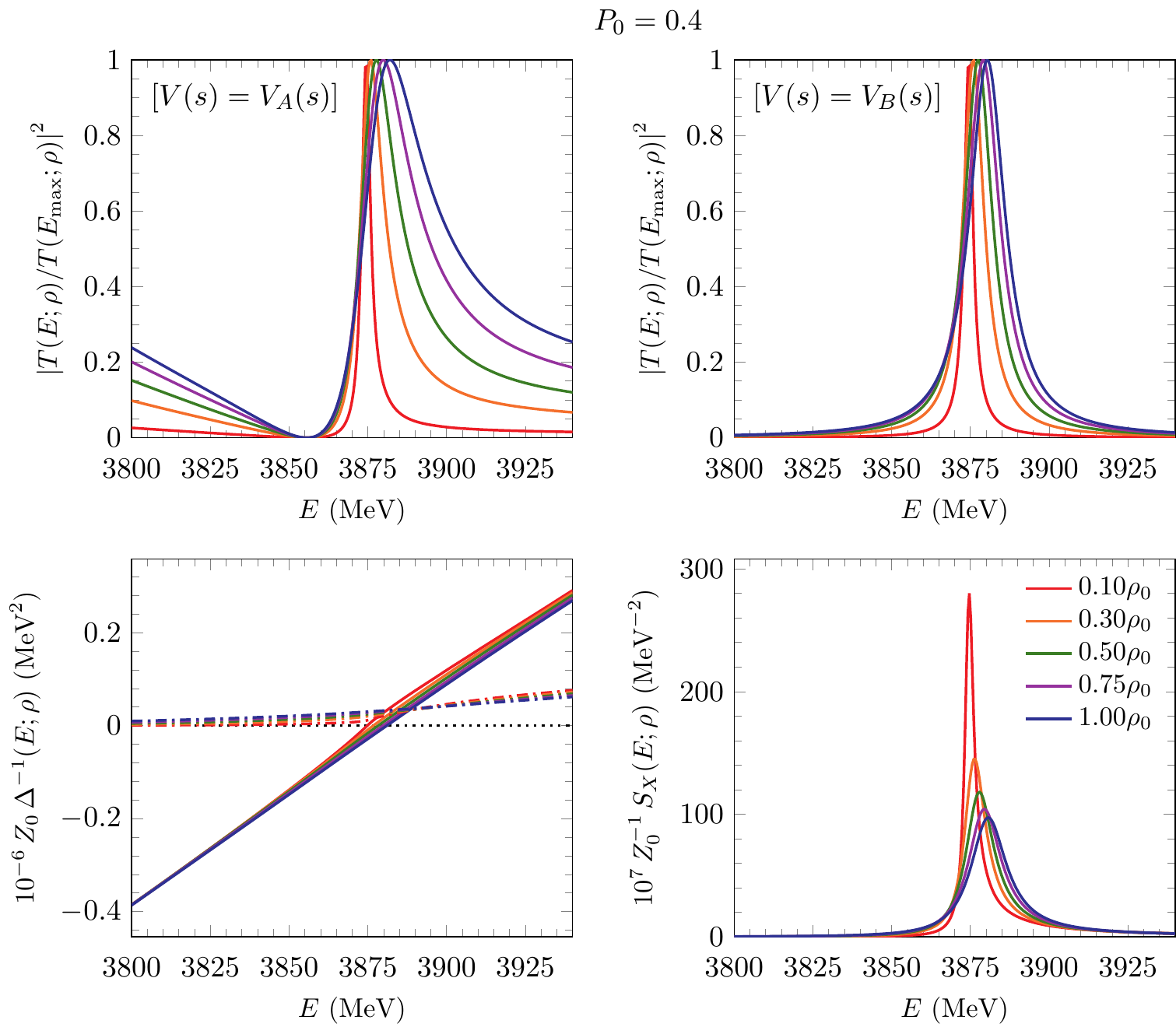}
\caption{\label{fig:PotAB_1} Top plots: Squared modulus of the amplitude $T(E; \rho)$, normalized to be one at the maximum, $E_\text{max}$, as a function of the center-of-mass energy of the $D\bar D^*$ pair, for a vacuum molecular probability $P_0=0.4$. The amplitudes are used computed using Eq.~\eqref{eq:Trho_basic} and the potentials $V_A(s)$ in Eq.~\eqref{eqs:V_pot_a} (left plot) or $V_B(s)$ in Eq.~\eqref{eqs:V_pot_b} (right plot). Bottom plots: Inverse of the propagator $\Delta(E; \rho)$ (left) and the spectral function $S_X(E;\rho)$ (right) for $P_0=0.4$, and multiplied by $Z_0=(1-P_0)$ and $Z_0^{-1}$, respectively. Neither the propagator, nor the spectral function  depend on the kernel $V(s)$, since they are determined by the vacuum $X(3872)$ and the in-medium two-meson loop function $\Sigma(E; \rho)$.}
\end{figure*}
\begin{figure*}\centering
\includegraphics[width=0.80\textwidth]{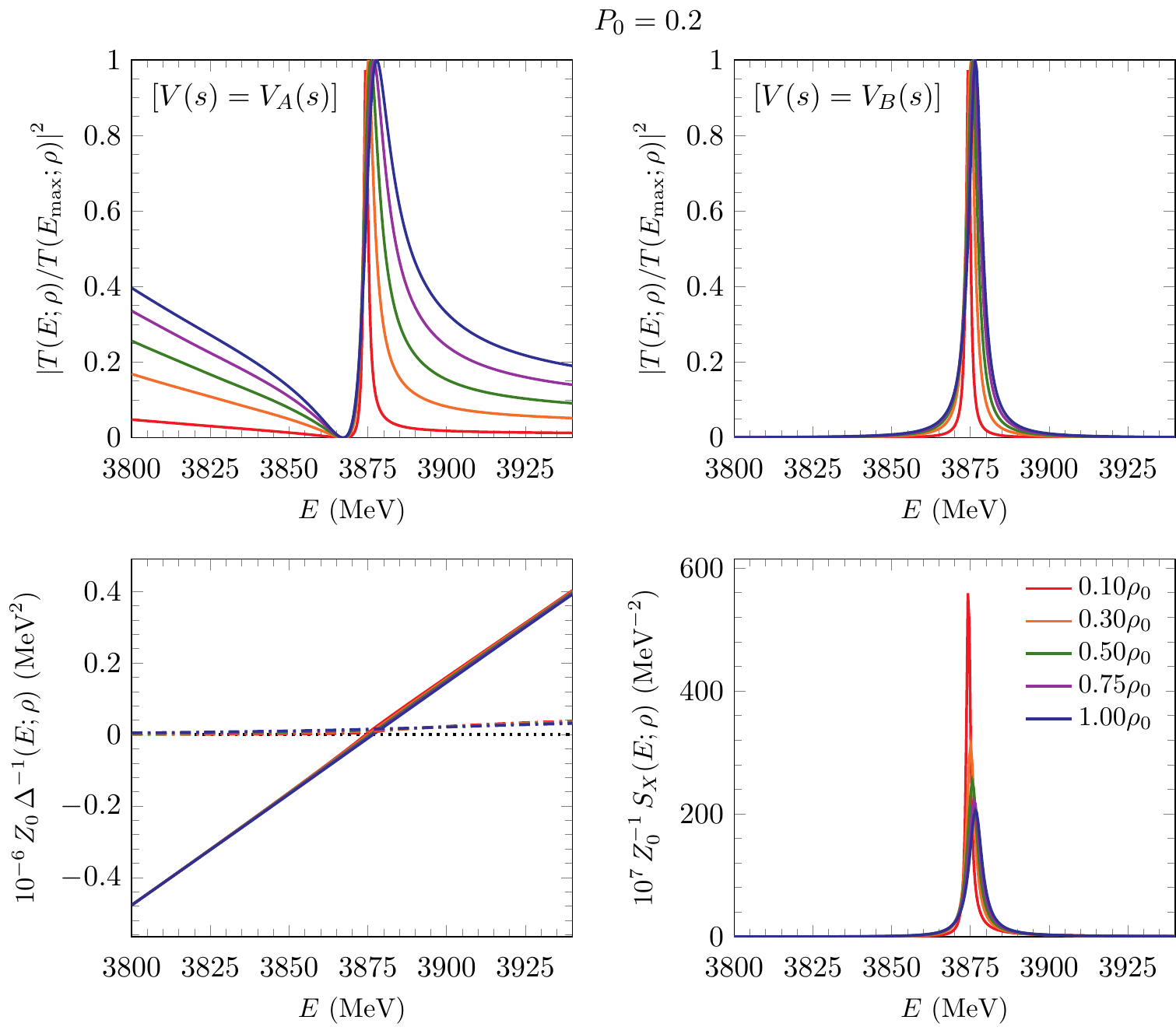}
\caption{\label{fig:PotAB_2} Same as Fig.~\ref{fig:PotAB_1}, but for $P_0=0.2$.}
\end{figure*}
Next, in Figs.~\ref{fig:PotAB_1} and \ref{fig:PotAB_2} we consider smaller molecular components, $P_0=0.4$ and $P_0=0.2$. As we discussed in Fig.~\ref{fig:sigmarho},  for these probabilities, the $V_A(s)$ [Eq.~\eqref{eqs:V_pot_a}] and $V_B(s)$ [Eq.~\eqref{eqs:V_pot_b}] potentials, despite leading to the same mass ($m_0$) and $D\bar D^*$ coupling ($g_0$) of the $X(3872)$ in the free space, considerably differ in the region of interest for the present  study. Hence, the corresponding $T-$matrices are different, even in the free space. Those deduced from $V_A$ show the zero that this potential has below $m_0$. As the molecular probability decreases, this zero gets closer to the $X(3872)$ vacuum mass, since the slope of $V_A(s)$ grows (in absolute value) as $1/P_0$. The position of the zero is independent of the nuclear density, being, however,  the dependence of the amplitude on the density clearly visible, both for energies below and above the energy, $E_0$, for which the potential and scattering amplitude vanish. Density effects for energies lower (higher) than $E_0$ become more (less) relevant for the $P_0=0.2$ case than for the $P_0=0.4$ one. 

In sharp contrast to the results stemming from $V_A(s)$, when $|T(E; \rho)|^2$ is computed using the $V_B(s)$ interaction, we see little structure beyond the peak induced by the bare pole present in the potential. The effects due to the medium dressing are small for $P_0=0.4$ and already quite difficult to disentangle for $P_0=0.2$. Hence, experimental input on $|T(E; \rho)|^2$, especially for energies below $E_0$, might shed light into the dynamics of the interacting $D\bar D^*$ pair that could be difficult to infer from their scattering in the free space.     

In Figs.~\ref{fig:PotAB_1} and \ref{fig:PotAB_2} we also show the inverse propagator $\Delta^{-1}(E;\rho)$ and the spectral function $S_X(E;\rho)$. These quantities do not depend on the type of potential employed --$V_A(s)$ or $V_B(s)$--, since they are determined by the vacuum $X(3872)$ and the in-medium two-meson loop function $\Sigma(E; \rho)$ given in Fig.~\ref{fig:sigmarho}. In what respects to the Im$[\Delta^{-1}(E;\rho)]$, the results here are the same as those discussed above in Fig~\ref{fig:ResT}, scaled down by the corresponding factor $P_0/(1-P_0)$. On the other hand, the plots for real part of $\Delta^{-1}(E;\rho)$ show that, for small molecular components, there is always a quasi-particle solution very close to $m_0$, and very little affected by the nuclear matter density. Finally, the spectral function $S_X(E; \rho)$ embodies the main features of $|T(E; \rho)|^2$ when the potential $V_B$ is used. However, it does not account for the medium modifications observed in the $T-$ matrix below $E_0$ when $V_A$ is employed.
\begin{figure*}\centering
\includegraphics{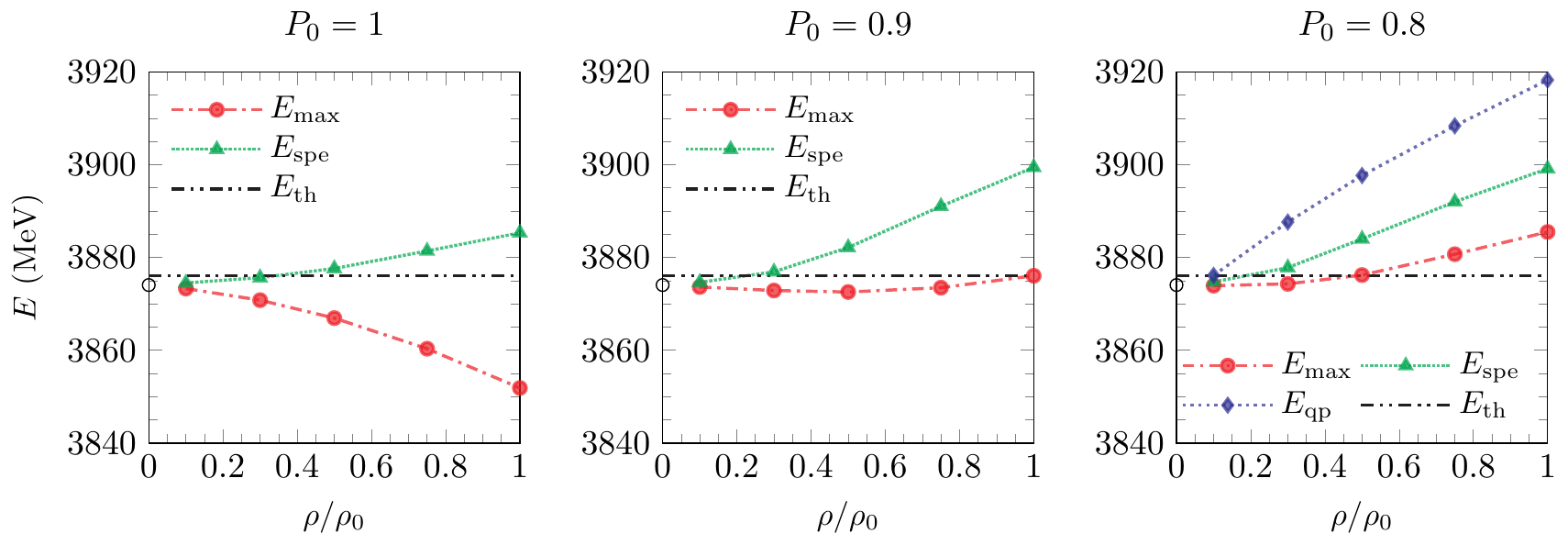}
\caption{\label{fig:EnergiesHighP} Positions $E_\text{max}$ and  $E_\text{spe}$ of the maxima of $|T(E; \rho)|$ and $S_X(E;\rho)$, respectively,  as a function of the nuclear matter density.  From left to right, the three plots show the cases $P_0=1$, $0.9$, and $0.8$. In the latter case, we also give the quasi-particle energy, $E_\text{qp}$, obtained by solving  Re$[\Delta^{-1}(E;\rho)]=0$. For these high molecular probabilities, the numerical differences due to the use of $V_A(s)$ or of $V_B(s)$ potentials [Eqs.~\eqref{eqs:V_pot_a} and \eqref{eqs:V_pot_b}, respectively] are very small. The black dashed-double dotted line represents the vacuum $D\bar D^*$ threshold, whereas the empty circle at $\rho=0$ is the $X(3872)$ vacuum mass $m_0$ of the $X(3872)$.}
\end{figure*}

\begin{figure*}\centering
\includegraphics{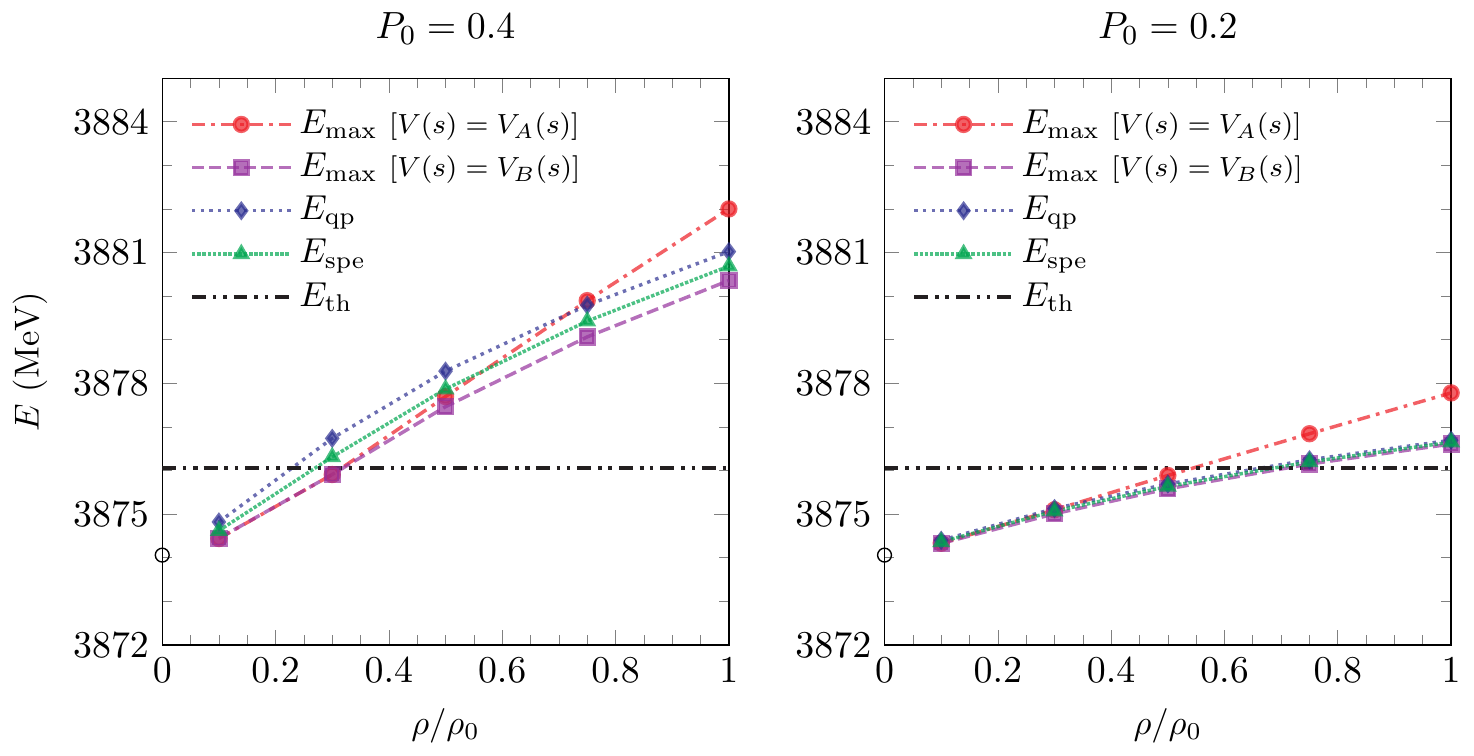}
\caption{\label{fig:EnergiesLowwP} Same as Fig.~\ref{fig:EnergiesHighP}, but for small molecular components, $P_0=0.4$ and $0.2$. We show separately $E_\text{max}$ obtained from $V(s)=V_A(s)$ or $V(s)=V_B(s)$ [Eqs.~\eqref{eqs:V_pot_a} and \eqref{eqs:V_pot_b}, respectively].}
\end{figure*}

To conclude, in Figs.~\ref{fig:EnergiesHighP} and \ref{fig:EnergiesLowwP}, we  show the positions $E_\text{max}$ and  $E_\text{spe}$ of the maxima of $|T(E; \rho)|^2$ and $S_X(E;\rho)$, respectively, for all molecular probabilities considered above in Figs.~\ref{fig:ResT}--\ref{fig:PotAB_2}. We also give the quasi-particle energies, $E_\text{qp}$, obtained by solving  Re$[\Delta^{-1}(E;\rho)]=0$ when they exist. For low values of $P_0$ (Fig.~\ref{fig:EnergiesLowwP}), we provide separately $E_\text{max}$ obtained from $V_A(s)$ or of $V_B(s)$ potentials. The results in these two figures reinforce the conclusions previously outlined. Indeed, we graphically see for the highest values of $P_0$, the appreciable difference between $E_\text{max}$ and  $E_\text{spe}$, with even an opposite density slope in the $P_0=1$ case. In Fig.~\ref{fig:EnergiesHighP}, we only observe for $P_0=0.8$ some resemblances between the maxima of $|T(E; \rho)|^2$ and $S_X(E;\rho)$, with  
quasi-particle energies well separated from both of them and exhibiting a significantly larger sensitivity with density. Medium effects are much smaller in Fig.~\ref{fig:EnergiesLowwP}, where results for $P_0=0.2$ and $P_0=0.4$ are collected. Some differences between $E_\text{max}$ obtained from $V_A$ or $V_B$ potentials are visible, even for the lowest of the molecular probabilities, for densities close to $\rho_0$. The quasi-particle and spectral-function energies are closer, and for $P_0=0.2 $ become indistinguishable from $E_\text{max}$ computed using $V_B$. This supports that, in this case, one is dealing with a compact state little affected by the dressing of the meson loops in the medium.

\subsection{Poles in the complex plane}\label{subsec:poles}

As already mentioned, the integral representation of Eq.~\eqref{eq:Sigmarho_basic} for the in-medium loop function $\Sigma(s;\rho)$ is not well suited for its continuation into the whole complex plane. The rich dynamical structure of the spectral functions $S_{D^{(*)}}$ and $S_{\bar{D}^{(*)}}$ shown in Fig.~\ref{fig:SEDmesons} is washed out by the  $\Omega$ and $\omega$ integrations implicit in Eq.~\eqref{eq:Sigmarho_basic} (see also Eq.~\eqref{eqs:Gifuns}). Thus,
almost no trace of the several peaks present in Fig.~\ref{fig:SEDmesons} can be distinctly appreciated in the resulting loop functions $\Sigma(s,\rho)$ depicted in Fig.~\ref{fig:sigmarho} for several densities. Actually, the latter are essentially equivalent to the loop function of a two-meson system regulated via a hard cutoff $\Lambda$, but evaluated with complex masses. Hence, we make the following approximation:
\newcommand{\Geff}{G^\text{(eff)}}
\newcommand{\comgloopappr}{\Sigma(s;\rho) \simeq \Geff(s;\rho) \equiv G(s,m_D^\text{(eff)}(\rho),m_{D^*}^\text{(eff)}(\rho))}
\begin{equation}\label{eq:effective_loop_G}
    \comgloopappr~,
\end{equation}
with $m_{D^{(*)}}^\text{(eff)}$ complex valued, and the superscript ``(eff)'' is included to remark that these are density-dependent effective masses, and do not correspond to the pole positions associated to the $D^{(*)}$ and $\bar D^{(*)}$ peaks in Fig.~\ref{fig:SEDmesons}. Additional details, including a discussion on the accuracy of the approximation,  can be found  in Appendix~\ref{app}. 

By means of the approximation in Eq.~\eqref{eq:effective_loop_G} we can now compute the in-medium $T_{D\bar D^*}(s;\rho)$ in the whole complex plane, for the different medium densities $\rho$ and vacuum probabilities $P_0$, and search for poles in the complex plane. We find a pole on the first Riemann sheet of the amplitude (as defined in Appendix \ref{app}), off the real axis. This does not represent any violation of the analyticity properties of the  complete-system scattering $T-$matrix, because of the effective procedure used to take into account  the  many body channels of the type $D\bar D^*N\to D\bar D^*N'$. In the present scheme, they are not explicitly considered in the coupled-channel space and only their effects on  $D\bar D^*\to D\bar D^*$  are included through the in-medium charmed-meson self-energies. 

\begin{figure}\centering
\includegraphics[height=8cm,keepaspectratio]{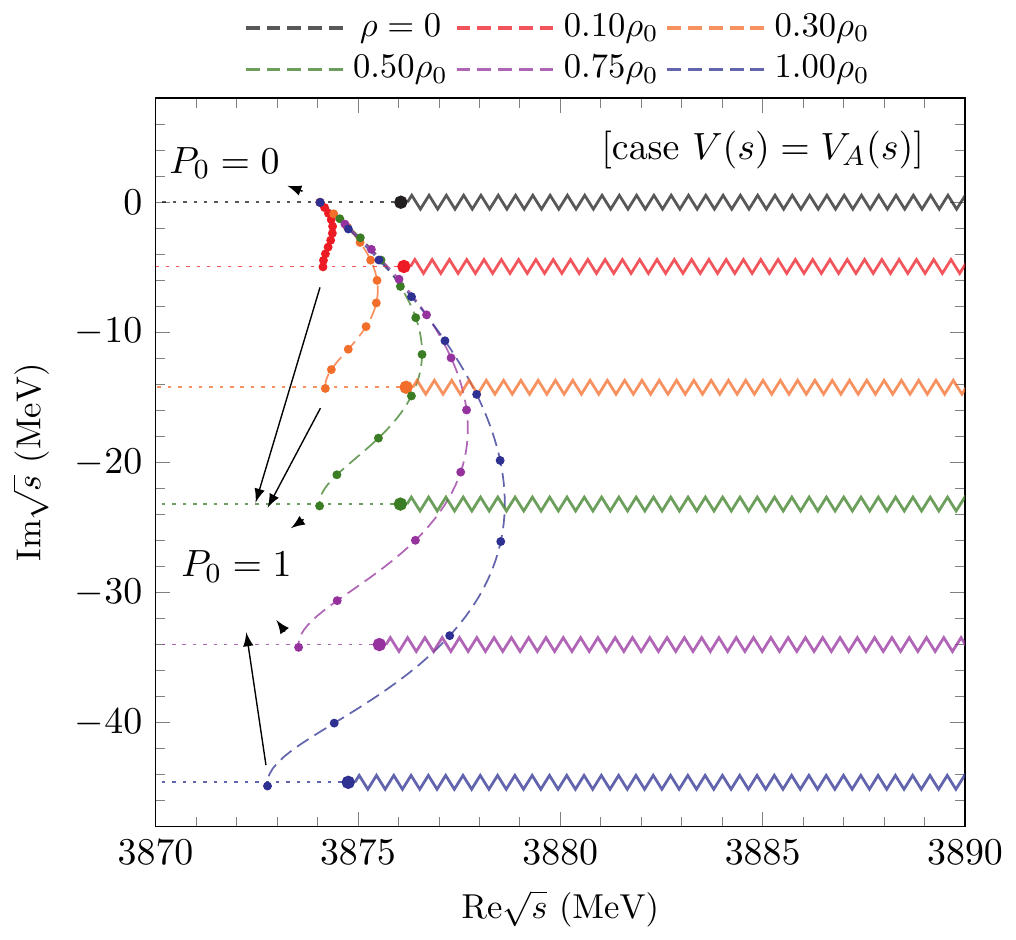}%
\includegraphics[height=8cm,keepaspectratio]{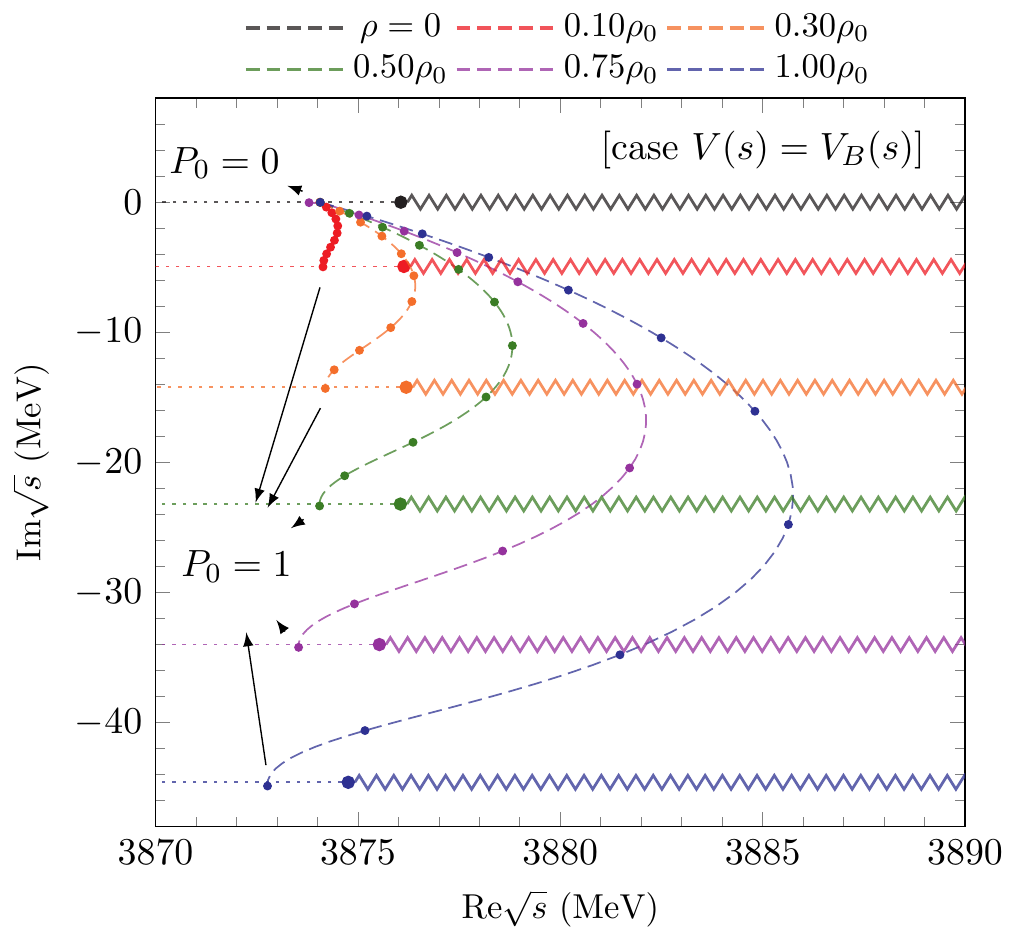}%
\caption{Complex pole position of the $X(3872)$ for different nuclear densities ($\rho$) and vacuum molecular probabilities ($P_0$). Results in the left and right plots have  been obtained using  amplitudes computed with $V(s)=V_A(s)$ [{\it cf.} Eq.~\eqref{eqs:V_pot_a}] and  $V(s)=V_B(s)$ [{\it cf.} Eq.~\eqref{eqs:V_pot_b}], respectively. The dashed curves show the continuous variation of the pole position with $P_0$, and the points represent steps in the probability $\Delta P_0 =0.1$. Different colors correspond to different nuclear densities, as detailed in the legend of the plots. The zigzag lines stand for the cut of the $\Geff(s;\rho)$ function (see text and Appendix~\ref{app} for further details).\label{fig:poles_evolP}}
\end{figure}

The pole position depends on the nuclear medium density $\rho$ and on the value chosen for the parameter $P_0$, the $X(3872)$ molecular probability in the vacuum. The pole position is represented in Fig.~\ref{fig:poles_evolP} for different values of $P_0$ and $\rho$, with each of  the colors associated to a particular density, and both $V_A(s)$ (left) and $V_B(s)$ (right) free space $D\bar D^*-$potentials considered in this work. For each density, the zigzag lines represent the loop function $\Geff(s;\rho)$ right hand cut:  
\begin{equation}
  \sqrt{s} \in \mathbb C \Big/ \left[\text{Im}\,p^2(s,m_{D^{(*)}}^\text{(eff)},m_{D}^\text{(eff)})=0\right] \, {\rm and} \, \left[\text{Re}\,p^2(s,m_{D^{(*)}}^\text{(eff)},m_{D}^\text{(eff)})>0\right]\,\label{eq:rhcut}
\end{equation}
extending to the right and starting at the branch point, $\sqrt{s}= (m_{D^{(*)}}^\text{(eff)}+m_{D}^\text{(eff)})$, where $p^2(s)=0$. In addition, $p(s)$ is defined in the Appendix. The dotted lines extending to the left represent the segments in which $\text{Im}\,p^2(s)=0$ and $\text{Re}\,p^2(s)<0$, where the density-dependent loop functions are thus real,\footnote{Because of the limited range in $\text{Re}\,\sqrt{s}$ explored in Fig.~\ref{fig:poles_evolP}, the curves in which $\text{Im}\,p^2(s)=0$ (the zigzag and dotted lines) look like straight lines, parallel to the real axis, although in general they are not, and have some curvature.} $\text{Im}\,\Geff(s;\rho)=0$. The dashed lines show the continuous variation of the pole position with $P_0$, where the points represent steps in the probability $\Delta P_0 =0.1$. When $P_0 \to 0$, {\it i.e.}, when the $X(3872)$  molecular component tends to vanish, the coupling of the $X(3872)$ to the $D^* \bar D$ channel tends to zero, and therefore, in this case, the pole remains at the original position in vacuum,  independently of the nuclear density. On the other end, when $P_0 \to 1$, {\it i.e.}, when the $X(3872)$ tends to be a purely molecular state, the pole appears to the left of the effective complex threshold, exactly in the segment where $\text{Im}\,p^2(s)=0$. This happens because, in this limit, the derivative term of the kernel $V_A(s)$ [{\it cf.} Eq.~\eqref{eqs:V_pot_a}] vanishes, and $V_A(s)$ is just a real constant.\footnote{Note that, as previously discussed, there is little difference between the results obtained with $V_A(s)$ or $V_B(s)$ when $P_0$ is close to one. Therefore, the argument presented here with $V_A(s)$ can be readily applied to the case of $V_B(s)$.} Therefore the pole, solution of $[1- V_A(s) \Geff(s;\rho)=0]$, should also satisfy  $\text{Im}\, \Geff(s;\rho)=0$. We also see in Fig.~\ref{fig:poles_evolP} that the in-medium $X(3872)$ pole position satisfies $\left\lvert \text{Im} \sqrt{s_P} \right\rvert \leqslant \left\lvert \text{Im} \left(m_D^\text{(eff)} + m_{D^*}^\text{(eff)} \right) \right\rvert$, {\it i.e.}, the $X(3872)$ width is always smaller than the sum of the $D$ and $\bar D^*$ effective widths. One can say that the pole position is {\it dragged} by the effective threshold $(m_D^\text{(eff)} + m_{D^*}^\text{(eff)})$, and that the effect is large or small depending on whether the in-vacuum probability $P_0$ is close to $1$ or to $0$, respectively. We also observe some dependence of the pole position, which as expected grows as the molecular content $P_0$ deviates from 1, on the used  $D\bar D^*$ interaction in the free space, namely $V_A(s)$ (Eq.~\eqref{eqs:V_pot_a}, left plot of Fig.~\ref{fig:poles_evolP}) or $V_B(s)$ (Eq.~\eqref{eqs:V_pot_b}, right plot of Fig.~\ref{fig:poles_evolP}).

\begin{figure}\centering
\includegraphics[height=6.7cm,keepaspectratio]{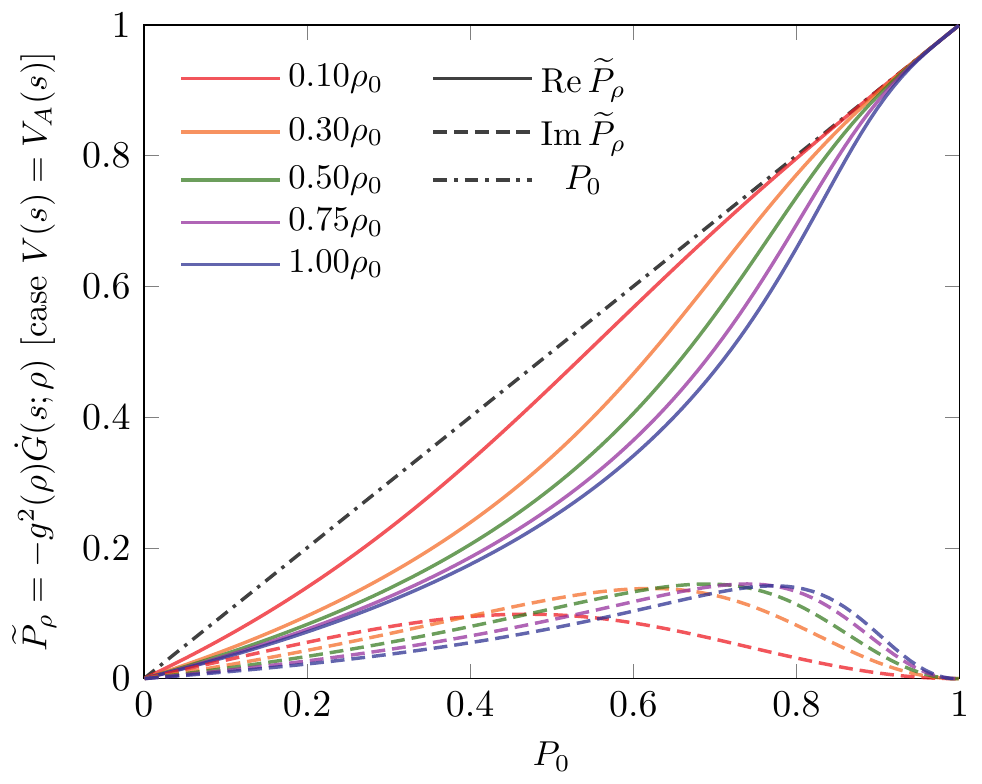}%
\includegraphics[height=6.7cm,keepaspectratio]{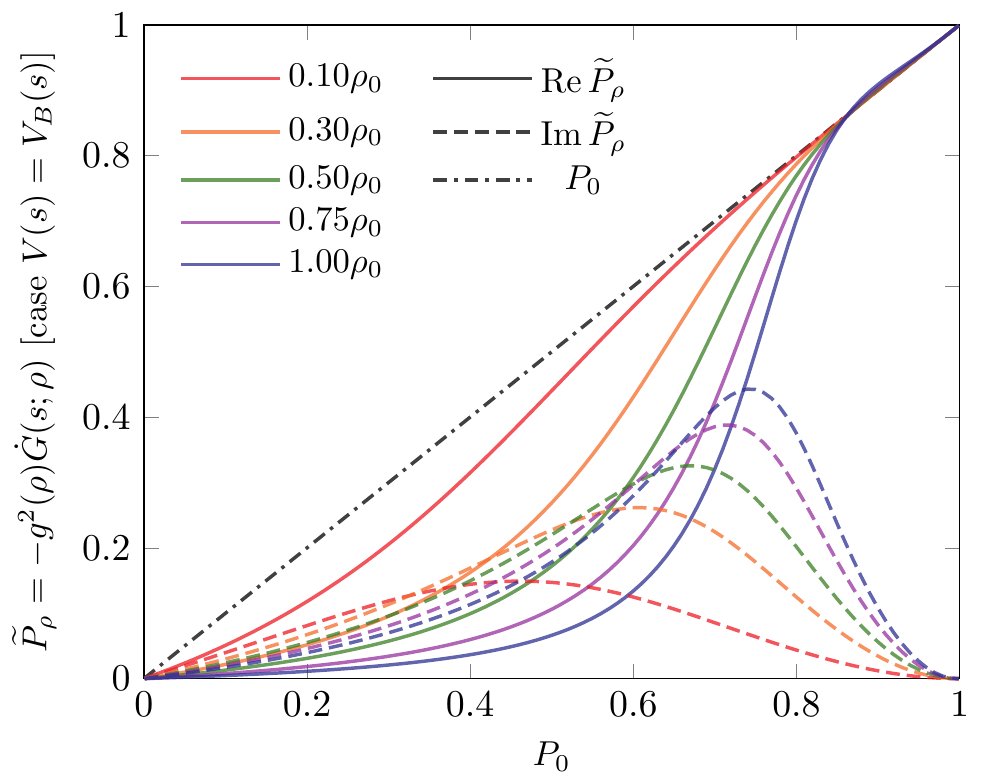}%
\caption{Dependence of the quantity $\widetilde{P}_\rho$ [{\it cf.} Eq.~\eqref{eq:prob_rho}] with the vacuum molecular probability $P_0$ for different densities. The solid (dashed) lines represent the real (imaginary) part of $\widetilde{P}_\rho$. The left and right plots correspond to the cases $V(s)=V_A(s)$ and $V(s) = V_B(s)$, respectively. \label{fig:prob_rho}}
\end{figure}

In our amplitudes, the vacuum molecular probability $P_0$ is a free parameter that we have varied to explore different scenarios. We can define the quantity $\widetilde{P}_\rho$,
\begin{equation}\label{eq:prob_rho}
\widetilde{P}_\rho = - g^2(\rho) \left. \frac{\mathrm{d} \Geff(s;\rho)}{\mathrm{d} s} \right\rvert_{s=m^2(\rho)}~,
\end{equation}
which generalizes to the nuclear medium the formula for the vacuum probability [{\it cf.} Eq.~\eqref{eq:Prob_vac}]. Since the pole position is in general complex, so will be this quantity. Therefore, in general, it will not be possible to interpret it as a probability.  In Fig.~\ref{fig:prob_rho}, we show $\widetilde{P}_\rho$  for different nuclear densities as a function of the vacuum probability $P_0$. This figure complements the results of Fig.~\ref{fig:poles_evolP}. We observe for this magnitude some quantitative differences between the results obtained with $V_A(s)$ (Eq.~\eqref{eqs:V_pot_a}, left plot of Fig.~\ref{fig:prob_rho}) or $V_B(s)$ (Eq.~\eqref{eqs:V_pot_b}, right plot of Fig.~\ref{fig:prob_rho}), but the qualitative behaviour is very similar. In the intermediate regions, far from the end points $P_0=0$ and $P_0=1$, the imaginary part of $\widetilde{P}_\rho$ can be sizeable, and for most of these values it increases with the density. In general, the effect of the nuclear medium in this intermediate $P_0$ region is to decrease both the real part and the modulus of $\widetilde P_\rho$ with respect to its original value $P_0$. However, we see that for both ends $P_0 \to 0$ or $P_0 \to 1$, we have that $\text{Im} \widetilde{P}_\rho \simeq 0$, and $\widetilde{P}_\rho \simeq P_0$. We thus see that in these cases the $X(3872)$ state can be said to conserve its original nature in the nuclear medium.

\section{Conclusions}
\label{sec:conclusions}

In this work we have studied the behaviour of the $\chi_{c1}(3872)$, also known as  $X(3872)$, in dense nuclear matter. The $X(3872)$ appears in the vacuum as a pole in the $D\bD^*$ scattering amplitudes, which are parametrized in a quite general form. The in-medium effects have been incorporated by dressing the $D \bD^*$ loop functions with the corresponding spectral functions of the charmed mesons. As a result, the $D \bar D^*$ amplitudes, when the charmed mesons are embedded in the nuclear medium,  have been determined for energies around the nominal $X(3872)$ mass. The $X(3872)$ spectral function has been also obtained for densities ranging up to  that of nuclear matter saturation.

For the kernel of the $D \bD^*$ scattering, we have used two possible energy-dependent potentials, each of them depending on two free parameters. Imposing that the vacuum amplitude has a pole in the physical Riemann sheet, these two parameters allow to fix the nominal $X(3872)$ mass and its coupling to the $D\bD^\ast$ channel, or, alternatively, the mass and the molecular probability $P_0$. Therefore, both types of interactions allow for the study of the $X(3872)$ as either a pure hadron-molecule state or a genuine quark state, as well as intermediate possibilities, in terms of $P_0$. However, both types of interactions have different analytical properties, which can give rise to different scattering amplitudes at finite density.

Using these two models for the interaction, we have explored the connection between the in-medium behaviour of the $X(3872)$ and its nature. In the case of the $X(3872)$ being mostly a molecular state, both interaction potentials behave similarly and lead to equivalent results for the in-medium  amplitudes. In this case, we have found that the $D\bar D^*$ amplitudes strongly depend on the density. The width of the $X(3872)$-peak significantly grows when the density is increased, while its position  moves to higher energies, as the molecular component is lowered. The $X(3872)$ spectral function follows the imaginary part of the $X(3872)$ self-energy, that increases with density due to the appearance of new many-body decay channels in matter. On the other hand, when smaller molecular components are considered, the $D \bar D^* $ amplitudes depend on the choice of the energy-dependent potential, specially for energies below the  free-space $X(3872)$ mass. Hence, the experimental input on the amplitudes at finite density might shed light into the dynamics of the $D \bar D^*$ interaction in the case of a state with a large genuine constituent quark component. Moreover, in this case, the $X(3872)$ spectral function, which is independent of the potential employed, is very little affected by the density.

The in-medium $D\bar D^*$ loop functions strongly depend on the interaction of $D$, $D^*$, $\bar D$ and $\bar D^*$ with nuclear matter. 
However, one can reasonably approximate them by a standard loop function evaluated with complex, effective masses of the $D^{(*)}$ and $\bD^{(*)}$ mesons. This fact allows for an analytical continuation of the loop function, and hence of the scattering amplitude, to the whole complex plane and to the second Riemann sheet. In turn, this allows for the search of the pole associated to the $X(3872)$ in the nuclear medium.  For finite density, the pole is found in the first Riemann sheet, but in the complex energy plane. However, this does not represent any violation of the analyticity properties of the $T-$matrix, because, in the present scheme, the $D\bar D^*N\to D\bar D^*N'$ many-body channels are not explicitly considered in the coupled-channel space, since their effects on  $D\bar D^*\to D\bar D^*$  are included via the in-medium charmed-meson self-energies. The behaviour of the $X(3872)$ pole with density is moreover fully in line with the change in matter of the squared modulus of the $T-$matrix amplitudes for real energies. Complex poles for the $X(3872)$ produced inside of a nuclear medium are collected in Fig.~\ref{fig:poles_evolP}, for different densities and free-space molecular probabilities. In the light of these results, we conclude that for the nuclear matter saturation density and molecular components of  the order of 60\% for the $X(3872)$, the many-body modes considered in this work provide widths for this resonance of around 30-40 MeV, and more modest mass-shifts (repulsive) with a maximum of 10 MeV. This latter outcome  contradicts the results obtained in the QCD-sum-rule calculation carried out in Ref.~\cite{Azizi:2017ubq} and based on a diquark-antidiquark picture for the $X(3872)$. Indeed, in the approach of Ref.~\cite{Azizi:2017ubq}, the mass-shift due to the nuclear matter is negative and is about 25\% ($\sim$ 800-900 MeV) when the saturation density is used. Therefore, any experimental analyses on the in-medium properties of $X(3872)$ and comparison of those with the results of the present study can increase our knowledge of the $X(3872)$ and help us gain useful information on the not well-known structure of this exotic state.

In this work we have studied the contribution of the dominant $D \bar D^*$ channel to the $X(3872)$ dynamics. In the future, we aim at extending our calculation to a more realistic situation by incorporating also coupled channels involving hidden-charm mesons, such as $J/\psi\,\pi$. Also, the results presented in this manuscript are based on a specific model for the $D^{(*)}N$ and $\bar D^{(*)}N$  interactions, which determine the in-medium modifications of the $D\bar D^*$ loop functions. Different or more elaborate models for these amplitudes could also be employed in the formalism we have derived here. In any case, our results indicate a very different behaviour with density of the $D \bar D^*$ amplitudes and the $X(3872)$ spectral function depending on the nature of the $X(3872)$. Thus, experiments that can access the nuclear finite-density regime, such as HiCs like CBM or those with fixed nuclear targets such as $\bar p$-nuclei in PANDA, are necessary and complementary to the spectroscopic analyses so as to discern the nature of $X(3872)$.

\acknowledgments
We thank E. Oset for valuable discussions at an early stage of this project, and for a careful reading of the manuscript. M.A. work supported by the U.S. Department of Energy, Office of Science, Office of Nuclear Physics under contract DE-AC05-06OR23177. L.T. acknowledges support from the Deutsche Forschungsgemeinschaft (DFG, German research Foundation) under the Project Nr. 411563442 (Hot Heavy Mesons), the CRC-TR 211 'Strong-interaction matter under extreme conditions'- project Nr. 315477589 - TRR 211, and the THOR COST Action CA15213. This research has been also supported  by the Spanish Ministerio de Econom\'ia y Competitividad, Ministerio de Ciencia e Innovaci\'on and the European Regional
Development Fund (ERDF) under contracts FPA2016-81114-P,  FIS2017-84038-C2-1-P, PID2019-105439G-C22 and
PID2019-110165GB-I00, by Generalitat Valenciana under contract PROMETEO/2020/023  and by the EU STRONG-2020 project under the program  H2020-INFRAIA-2018-1, grant agreement no. 824093.

\appendix

\section{\boldmath Further details on $\Geff(s;\rho)$}\label{app}
\begin{figure}\centering
\includegraphics[scale=0.9]{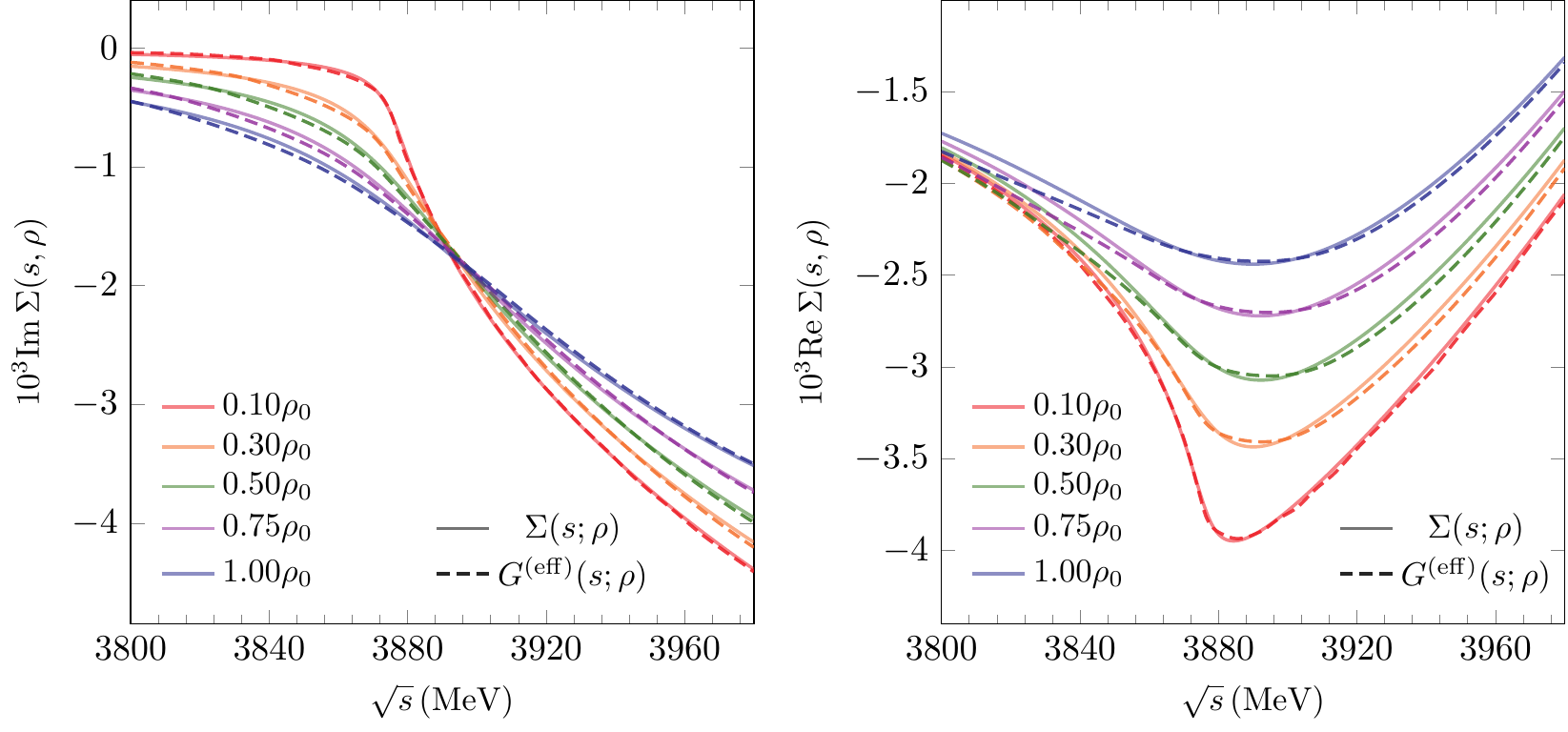}
\caption{The original loop function $\Sigma(s;\rho)$ (solid lines), shown in Fig.~\ref{fig:sigmarho}, compared with the approximated one, $\Geff(s;\rho)$, obtained from Eq.~\eqref{eq:effective_loop_G} (dashed lines). The imaginary and real parts of both functions, as a function of the c.m. energy of the $D\bar D^*$ pair are displayed in the left and right plots, respectively.\label{fig:Geff}}
\end{figure}

In this Appendix we give further details on the approximation made in Sec.~\ref{subsec:poles}, and on the analytical properties of the loop function employed. The approximation is:
\begin{equation*}
    \comgloopappr~,\tag{\ref{eq:effective_loop_G}}
\end{equation*}
where  $G(s,m_1,m_2)$ can be computed using the explicit formulas given for instance in Ref.~\cite{Oller:1998hw} (see in particular the {\it erratum}), regulated with a momentum cutoff of $0.7\ \text{GeV}$. In addition, we take for the density dependent effective masses
\begin{subequations}\label{eqs:effective_loop_masses}
\begin{align}
    m_D^\text{(eff)}(\rho) & = m_D + \Delta m(\rho) - i \frac{\Gamma(\rho)}{2}~,\\
    m_{D^*}^\text{(eff)}(\rho) & = m_{D^*} + \Delta m(\rho) - i \frac{\Gamma(\rho)}{2}~.
\end{align}
\end{subequations}
with $m_{D^{(*)}}$, the vacuum masses, and $\Delta m(\rho)$ and $\Gamma(\rho)$ real quantities. We note that in the $m_{D}^\text{(eff)}$ and $m_{D^{*}}^\text{(eff)}$ definitions we have forced a common shift $\Delta m(\rho) - i \frac{\Gamma(\rho)}{2}$ with respect to the vacuum masses. Being this an effective representation, we find that this {\it ansatz} is enough to approximate the original loop function, $\Sigma(s,\rho)$. In Fig.~\ref{fig:Geff} we show in the left (right) panel the imaginary (real) part of the loop function $\Sigma(s,\rho)$ together with the approximation determined by Eq.~\eqref{eq:effective_loop_G}, computed with the parameters $\Delta m(\rho)$ and $\Gamma(\rho)$ collected in Table~\ref{tab:effectivemasses}. The latter are chosen so as to approximately match the original loop functions $\Sigma(s;\rho)$ for the different densities considered in this work. As can be seen, the approximation works reasonably well.

\begin{table}[h]\centering
    \begin{tabular}{|c|r|r|} \hline
$\rho/\rho_0$ & $\Delta m(\rho)\ \text{(MeV)}$ & $\displaystyle -\frac{\Gamma(\rho)}{2}\ \text{(MeV)}$ \\ \hline
$0.10$ & $+0.04$ & $ -2.5$ \\
$0.30$ & $+0.07$ & $ -7.1$ \\
$0.50$ & $-0.01$ & $-11.6$ \\
$0.75$ & $-0.26$ & $-17.0$ \\
$1.00$ & $-0.65$ & $-22.3$ \\ \hline
    \end{tabular}
    \caption{Parameter values used to determine the effective masses $m_D^\text{(eff)}(\rho)$ and $m_{D^*}^\text{(eff)}(\rho)$ [Eqs.~\eqref{eqs:effective_loop_masses}] for different nuclear densities $\rho$.\label{tab:effectivemasses}}
\end{table}

\begin{figure}\centering
\includegraphics[scale=0.85]{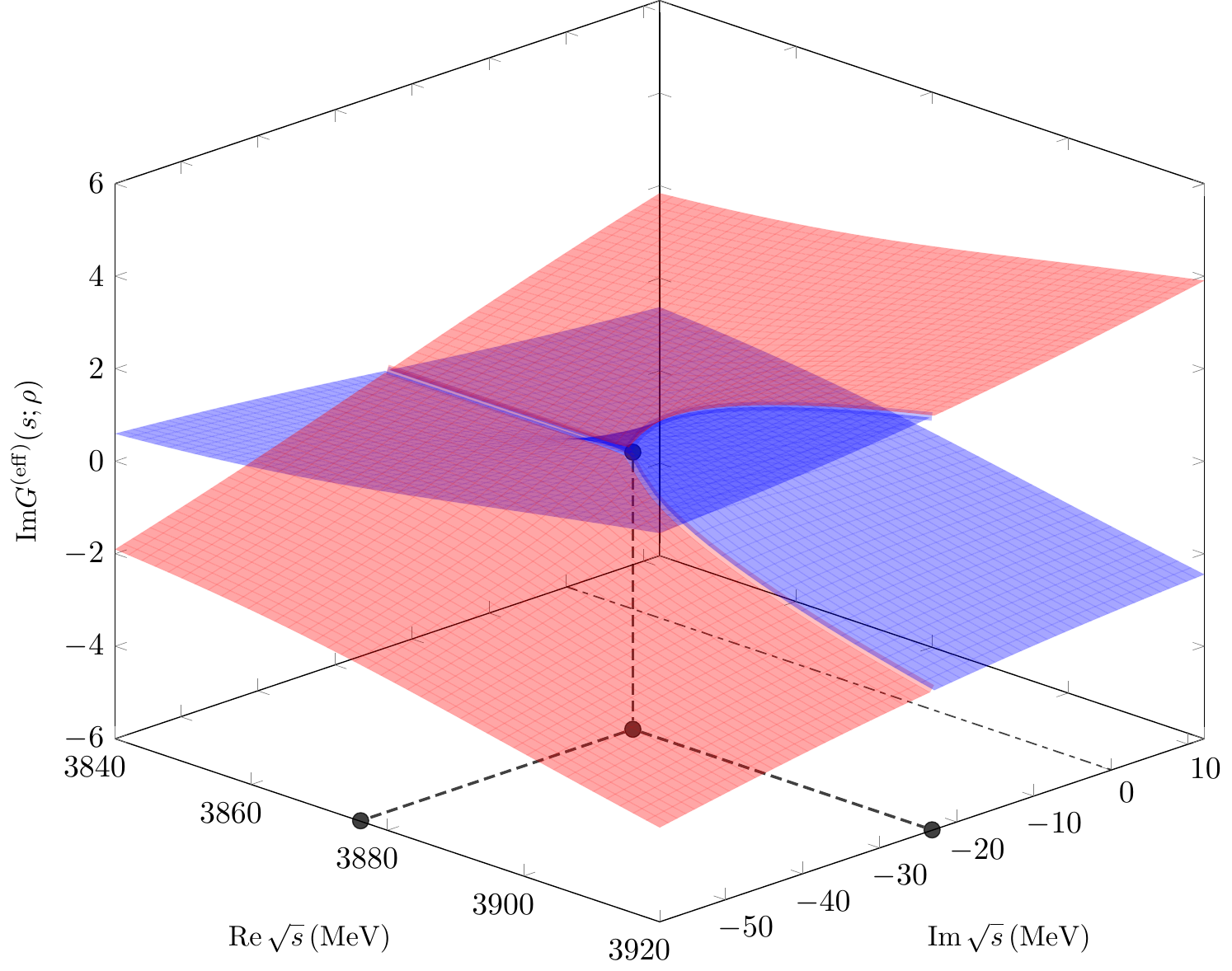}
\caption{The function $\Geff(s;\rho)$ for the case $\rho=\rho_0/2$ on the $\sqrt{s}-$complex plane. The first (second) Riemann sheet is shown in blue (red).\label{fig:G3D}}
\end{figure}

The loop function $\Geff(s;\rho)$ can be continued analytically to the whole complex plane, and the second (or nonphysical) Riemann sheet is defined as:
\begin{eqnarray}
    \Geff_\text{II}(s;\rho) &=& \Geff(s;\rho) + i \frac{p[s,m_D^\text{(eff)}(\rho),m_{D^*}^\text{(eff)}(\rho)]}{4\pi \sqrt{s}}\,,\nonumber \\ 
    p(s,m_1,m_2)&=& \frac{\left[s-(m_1+m_2)^2\right]^{\frac12}\left[s-(m_1-m_2)^2\right]^\frac12}{2\sqrt{s}} \label{eq:FRS-SRS}
\end{eqnarray}
In Fig.~\ref{fig:G3D} and for $\rho=\rho_0/2$, we show in blue (red) the physical (nonphysical) Riemann sheet of the function $\Geff(s;\rho)$ in the $\sqrt{s}-$complex plane. The cut of $\Geff(s;\rho)$ lies on a curve for the variable $\sqrt{s}$, given in Eq.~\eqref{eq:rhcut} of the main text, which in the free space ($\rho \to 0$) is the usual right hand cut, $\sqrt{s} > (m_D + m_{D^*})$ on the real axis, with $\sqrt{s}=(m_D+m_{D^*})$ the branch point. For finite density and therefore complex masses, this branch point moves from the real axis into the complex plane, and the cut does not lie in the real axis either.

\bibliographystyle{apsrev4-1_MOD}
\bibliography{X3872_medium.bib}

\end{document}